\pdfoutput=1
\documentclass[%
 reprint,
superscriptaddress,
 amsmath,amssymb,
 aps,
floatfix,
]{revtex4-1}

\usepackage{graphicx}
\usepackage{dcolumn}
\usepackage{bm}
\usepackage{color,soul}
\usepackage{float} 
\usepackage{subcaption,natbib}

\usepackage[above,below]{placeins}	

\begin{document}

\preprint{APS/123-QED}

\title{Characterizing active learning environments in physics using network analysis and COPUS observations}

\author{Kelley Commeford}
\author{Eric Brewe}%
\affiliation{%
 Drexel University\\
 3141 Chestnut St, Philadelphia, PA 19104%
}

\author{Adrienne Traxler}
\affiliation{
 Wright State University\\
 3640 Colonel Glenn Hwy, Dayton, OH 45435
}%

\date{\today}

\begin{abstract}
This study uses social network analysis and the Classroom Observation Protocol for Undergraduate STEM (COPUS) to characterize six research-based introductory physics curricula. Peer Instruction, Modeling Instruction, ISLE, SCALE-UP, Context-Rich Problems, and Tutorials in Introductory Physics were investigated. Students in each curriculum were given a survey at the beginning and end of term, asking them to self-identify peers with whom they had meaningful interactions in class. Every curriculum showed an increase in the average number of student connections from the beginning of term to the end of term, with the largest increase occurring in Modeling Instruction, SCALE-UP, and Context-Rich Problems. Modeling Instruction was the only curriculum with a drastic change in how tightly connected the student network was.  Transitivity  increased for all curricula except Peer Instruction. We also spent one week per research site in the middle of the term observing courses using COPUS. From these observations, the student COPUS profiles look nearly the same for Tutorials, ISLE recitations, and Context-Rich Problems discussion sections. This is likely due to the large resolution of activities that can be coded as ``other group activity,'' suggesting the need for a more detailed observation instrument.

\end{abstract}

\pacs{Valid PACS appear here}
\maketitle


\section{\label{sec:Motivation}Introduction}  

After decades of research in discipline based education research communities, it is well established that active learning is more effective than standard passive lecture at improving student outcomes \cite{Freeman, Hake1998}. A vast majority of active learning research still uses passive lecture methods as a baseline measurement of learning techniques. 
As such, \citet{Freeman} 
recommend a second wave of discipline based education research initiatives that study active learning methods independently of passive lecture methods, so we can understand the mechanisms through which active learning promotes increased student outcomes. 

However, studies of active learning pedagogies as independent entities have encountered difficulties because implementations vary widely. When comparing learning gains of individual implementations of Tutorials in Introductory Physics, with different instructors and student populations, there have been inconsistent gains based on instructor background and student buy-in~\cite{Pollock2008SustainingPhysics,FinkelsteinN.D.andPollock2005ReplicatingPhysics}. Additionally, a study of several Peer Instruction classrooms at the same university, but with different instructors, shows that the implementation of the same pedagogical methods vary greatly, resulting in differing classroom norms~\cite{Turpen2009NotInstruction}. Implementation of a given active learning pedagogy can also be severely limited based on external factors, such as assigned classroom, class size, and student preparation~\cite{Henderson2008}. Before we can even think of directly comparing pedagogies to each other, we need to understand the roles that different mechanisms play when implementing a pedagogy, and what impact those mechanisms have on students interacting with the curriculum.
 
To understand how active learning mechanisms impact student achievement, we first need to develop an appropriate vocabulary to describe the distinguishing features of an active learning environment. Doing so will allow researchers to discuss these implementations as individual entities, rather than as umbrella ``active learning methods" to be compared to passive lecture. 

Active learning environments vary based on student and instructor tasks and behaviors. There have been a handful of observational characterization studies done for Peer Instruction~\cite{Wood2016CharacterizingClass} and small group physics workshops~\cite{West2013VariationCourse}. This study aims to broaden that lens to include six active learning pedagogies commonly used in physics, and we propose to characterize these activities using two complementary methods, the Classroom Observation Protocol for Undergraduate STEM (COPUS)~\cite{Smith2013} and social network analysis.

Active learning, at its core, provides opportunities for students to interact with each other. COPUS can be used to record instructor and student interaction categories with the aim of creating a unique profile for each implementation. Social network analysis can be used as a means to quantify student interactions that arise during a given active learning implementation. There is evidence that a student's position in the classroom social network improves course outcomes and persistence in physics~\cite{Bruun2013TalkingScores,Zwolak2017}, so it is natural to surmise that active learning contributes to social network development.

With COPUS profiles and classroom social network data, we can begin describing how classroom tasks and behaviors can correlate to student social network development and mobility. By understanding what students are doing and how it impacts their social position within the class, future research can aim to better understand how different kinds of interactions lead to student growth. The goal of this paper is to describe the characteristics of six distinct pedagogies in physics, as measured by COPUS and network analysis. We do not aim to directly compare pedagogies.

We begin with a brief literature review to introduce the observation protocol, social network analysis concepts, and the active learning pedagogies that are studied. Then, we describe how observation sites were chosen, the demographic information of each site, and how data was collected. Finally, we show data from each observed pedagogy and discuss how the in-class tasks and behaviors present themselves in the social network data.  

\section{Literature Review}

\subsection{Observation Protocol}

A research-based pedagogy is generally accepted as `active learning' if the students are involved in the learning process in some meaningful way. Additionally, a pedagogy is considered `active learning' if it is based on research with regards to development and implementation, and students consistently show learning gains when the pedagogy is implemented in the classroom~\cite{Meltzer2012ResourcePhysics}. However, it is difficult to measure how much active learning is occurring in a classroom, largely due to the wide array of active learning methods that exist. The most straightforward way to measure `active learning-ness' is to conduct classroom observations using some sort of protocol. Several observation protocols exist that serve various purposes, some of which will be discussed briefly here. 

There are two categories of observation protocols: open-ended and structured~\cite{Smith2013}. Open-ended protocols typically provide the observer with prompting questions, to which they provide feedback. Structured protocols, on the other hand, provide the observer with some sort of framework to input observation data. Since we aim to have concrete measurements of how much and what kind of active learning is occurring, we discarded all protocols that were 100\% open-ended, as they rely on observer judgements and generally focus more on opinion statements rather than measurable quantities. A discussion of structured protocols that were considered for this study follows. 

The Reformed Teaching Observation Protocol (RTOP)~\cite{Sawada2002MeasuringProtocol} has observers make holistic judgements about the quality of lesson design and implementation, classroom culture, and content coverage. While it includes the use of Likert scales to assign numerical values to these categories, these categories are largely subjective. Additionally, the RTOP observation items do not fully capture the time-dependence of student and instructor activities and interactions. RTOP also has a long training time due to the in-depth theoretical framework that it is built upon. Finally, RTOP gives a numeric score where higher is better, which we wanted to avoid because our purpose is comparison and description rather than ranking. The RTOP may be promising for future studies delving into the quality of a given pedagogical implementation, assuming inter-rater reliability could be reached.

The Real-Time Instructor Observation Tool, or RIOT, is a computer-based tool that records instructor behaviors in real time~\cite{West2013VariationCourse}. This protocol gives a fine-grained temporal observation of instructor interactions with students. While this tool is valuable for understanding how instructors are leading conversations in the classroom, it was ultimately not chosen due to the computational hardware requirement and lack of student observation codes.  This protocol is best suited for studying instructor behaviors independently of student behaviors. 





The Teaching Dimensions Observation Protocol (TDOP)~\cite{Hora2008TeachingManual} has 46 codes to delineate student and instructor behaviors in the classroom, as well as a handful of open-ended responses. The observer codes interactions during a two minute time interval; if the behavior occurred for longer than five seconds, it gets coded. This protocol requires an extensive three-day training to achieve inter-rater reliability. While this protocol gives a broad overview of what is happening in the classroom, it was deemed impractical for our needs due to long training and large number of codes to keep track of during live observations.


The Classroom Observation Protocol for Undergraduate STEM (COPUS)~\cite{Smith2013} is similar in structure to the TDOP, but has 25 codes instead of 46, and does not have observation codes that require the observer to make judgements about the quality of instruction. It has been shown to have strong inter-rater reliability after a mere one and a half hour training period. The non-judgmental coding schemes provide a quantitative view of the classroom in two-minute intervals. Due to the small number of codes, it is well known for its ease of use in a live classroom observation~\cite{Smith2013}. 

We ultimately chose COPUS as the observation protocol for this study because we wanted to get an overall picture of \emph{what} was happening in the classroom, as opposed to how effective or well implemented the pedagogy was. In this study, we intentionally chose development sites or secondary sites recommended by the developers of each pedagogy to ensure high-fidelity implementations, so the `instruction quality' codes were deemed unnecessary for our initial investigation. Additionally, we want our investigations to be easily reproducible, so the shorter training and high inter-rater reliability were highly desirable, despite only having one rater for this study (and thus no inter-rater reliability demonstrations are presented). While COPUS is limited in its ability to measure code duration, as codes are recorded in two-minute intervals instead of instantaneously, it includes student and teacher behavior codes that are appropriate for examining active learning pedagogies. Codes included in the COPUS protocol can be seen in Table \ref{tab:COPUSCodes}. Additional protocols that address the time resolution concern were not appropriate or not published at the time the project started.

\begin{table*}[htbp]
  \caption{COPUS codes for instructor and student activities. We have summarized the descriptions for brevity, but the full code descriptions can be found in \citet{Smith2013}. \label{tab:COPUSCodes}}
  \begin{ruledtabular}
    \begin{tabular}{ll}
      \textbf{Students are Doing}& \\
      \hline \\
      \textbf{L} & \textbf{Listening} to instructor/taking notes, etc.  \\
      \textbf{Ind} & \textbf{Individual thinking/problem solving}. Marked when instructor explicitly asks students to think \\ &about a clicker question or another question/problem on their own  \\
      \textbf{CG} & \textbf{Clicker group} discussion with 2 or more students  \\
      \textbf{WG} & \textbf{Working in groups} on worksheet activity  \\
      \textbf{OG} & \textbf{Other group} activity, like a lab  \\
      \textbf{AnQ} & \textbf{Answering a question} posed by the instructor with rest of class listening  \\
      \textbf{SQ} & \textbf{Student asks question} to instructor \\
      \textbf{WC} & \textbf{Whole class} discussion \\
      \textbf{Prd} & \textbf{Predicting} the outcome of demo or experiment  \\
      \textbf{SP} & \textbf{Student presentation}  \\
      \textbf{TQ} & \textbf{Test or quiz}  \\
      \textbf{W} & \textbf{Waiting} (instructor late, working on fixing AV problems, instructor otherwise occupied, etc.)  \\
      \textbf{O} & \textbf{Other}  \\
      
     \hline\\
      \textbf{Instructor is Doing}& \\
      \hline\\
      \textbf{Lec} & \textbf{Lecturing} (presenting content, deriving mathematical results, presenting a problem solution, etc.)  \\
      \textbf{RtW} & \textbf{Real-time writing} on board, document projector, etc.\ (often checked off along with Lec)  \\
      \textbf{FUp} & \textbf{Follow-up}/feedback on clicker question or activity to entire class  \\
      \textbf{PQ} & \textbf{Posing questions} to students (non-clicker and non-rhetorical)  \\
      \textbf{CQ} & \textbf{Clicker question} (mark the entire time the instructor is using a clicker question, not just \\&when first asked)  \\
      \textbf{AnQ} & \textbf{Answering questions} from students with entire class listening  \\
      \textbf{MG} & \textbf{Moving and guiding} ongoing student work during active learning task \\
      \textbf{1o1} & \textbf{One-on-one} extended discussion with one or more individuals, not paying attention to the rest of \\&class (can be along with MG or AnQ)  \\
      \textbf{D/V} & \textbf{Demo, video}, experiment, simulation, or animation  \\
      \textbf{Adm} & \textbf{Administration} (assign homework, return tests, etc.)  \\
      \textbf{W} & \textbf{Waiting} when there is an opportunity for an instructor to be interacting with or observing/listening\\& to student or group activities and the instructor is not doing so  \\
      \textbf{O} & \textbf{Other}  \\
    \end{tabular}
  \end{ruledtabular}
\end{table*}

\subsection{Network Analysis}
Social network analysis has been used as an analysis tool in a Modeling Instruction classroom previously. Brewe {\em et.\ al.}\ showed that Modeling Instruction produced classroom networks that were structurally unique from a network formed in a standard passive lecture environment~\cite{Brewe2010ChangingCommunities}, which was a large motivation for this project. Meanwhile, Zwolak, Dou, and collaborators showed how social network analysis can be used to correlate student social positioning with other factors, such as persistence \cite{Zwolak2017} and self-efficacy \cite{Dou2016BeyondLens}. The survey methods deployed by Brewe {\em et.\ al.}\ were refined by Zwolak and Dou, which subsequently influenced the survey methods of this study. 

Social network analysis has also been done in the context of upper division physics courses, where homework grades were found to be strongly correlated with student centrality in their homework problem-solving network~\cite{Vargas2018CorrelationPerformance}. These centrality measures remained stable across different courses and types of assignments. 
\citet{Vargas2018CorrelationPerformance}
showed the potential benefits of studying physics classrooms from a network perspective to understand how student gains manifest.

We ultimately chose network analysis as a tool to study active learning environments due to its applicability to relational data, and its successful implementation in a physics context, as demonstrated independently by 
the studies above \citep{Zwolak2017,Dou2016BeyondLens,Brewe2010ChangingCommunities,Vargas2018CorrelationPerformance}. 
We expand on these ideas by collecting social network data from six pedagogies to identify the network structures that arise when a given pedagogy is used. Before we begin, we need to take a moment to define some network terms. 

\begin{figure}[h]
    \includegraphics[width=\linewidth]{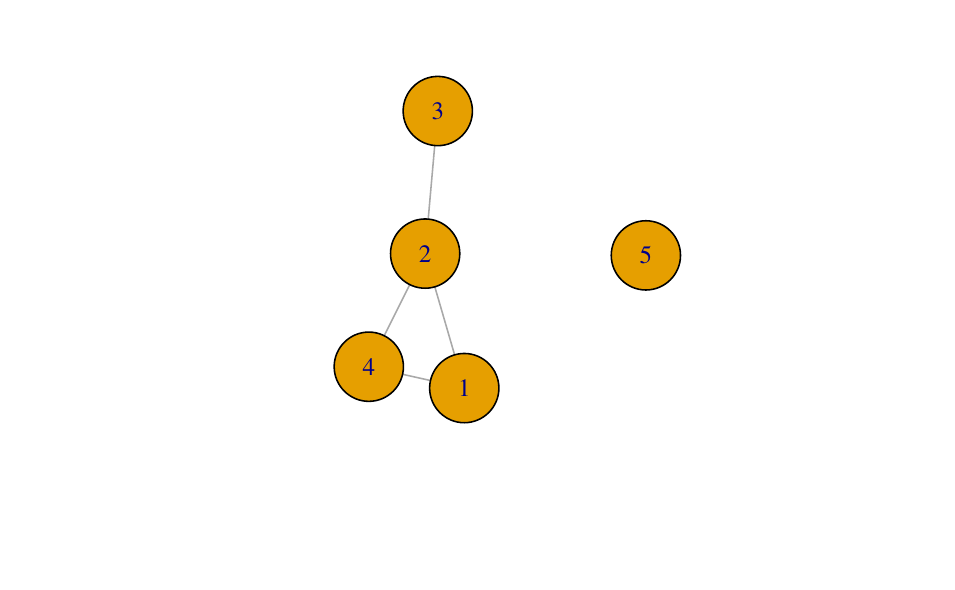}
    \caption{An undirected toy network, for illustration purposes only. If an interaction occurred, the edge exists regardless of which node initiated the interaction. }
    \label{fig:toynetwork-undirected}
\end{figure}

We have provided a toy network to help illustrate these terms in figure \ref{fig:toynetwork-undirected}. We use undirected networks in this study, which means ties exist regardless of which student indicated contact. The following terms are defined in the context of undirected networks.
\\


\noindent\textbf{Node} -- Sometimes referred to as an actor, the node is the noun in the network. In our toy network, the nodes are represented by the orange dots. In this study, the nodes represent the students in the class.
\\
\\
\textbf{Edge} -- Sometimes referred to as a tie, an edge is the verb in the network. Edges are represented by the lines connecting the nodes in the toy network diagrams. In this study, the edges indicate communication between students.
\\
\\
\textbf{Degree} -- Degree measures how many edges a given node is connected with. For example, in figure \ref{fig:toynetwork-undirected}, node 1 is connected to nodes 2 and 4, for a total of two connections. Node 1 has a degree of 2. 
\\
\\
\textbf{Density} -- The density is the ratio of actual number of edges in a network to possible number of edges. In our undirected toy network (Fig.\ \ref{fig:toynetwork-undirected}), there are 4 edges, but 10 possible unique edges, giving us a density of 2/5. 
\\
\\
\textbf{Diameter} -- If we calculate the geodesic distance (shortest path) between any two nodes, and then do this for every pair of nodes in the network, the longest of those paths is the diameter of the network. In our toy network (Fig.\ \ref{fig:toynetwork-undirected}), our longest-shortest path is from 3$\rightarrow$1 via the path 3$\rightarrow$2$\rightarrow$1, for a diameter of 2.
\\
\\
\textbf{Transitivity} -- In an undirected network, this is the ratio of closed triplets to triads. A closed triplet in an undirected network looks like the arrangement of nodes 1$\rightarrow$2$\rightarrow$4, while a triad would be one step smaller and missing the closure, like 1$\rightarrow$2$\rightarrow$3.  High levels of transitivity are usually indicative of a collaborative environment, so this measure will be of particular importance when considering network formation in active learning environments. 
\\
\\
\textbf{Giant component} -- The giant component is the number of nodes that are all connected in the largest cluster of nodes. In our toy model, our largest cluster has 4 nodes in it, so the giant component would be 4.

\subsection{Active Learning Pedagogies}

This project investigated six active learning pedagogies commonly used in introductory physics courses at the university level. These pedagogies have been featured in the New Faculty Workshops run by the American Association of Physics Teachers~\cite{Henderson2008} and have research articles describing their development, implementation, and outcomes (and can thus be considered active learning according to \citet{Meltzer2012ResourcePhysics}). 

\subsubsection{Tutorials in Introductory Physics}
Tutorials in Introductory Physics have been iteratively researched and developed at the University of Washington~\cite{McDermott2002}. This curriculum is typically implemented in a traditional lecture/lab/recitation setup, with the bulk of the tutorial material presented in recitation (referred to as the tutorial section). Tutorials focus on building a strong conceptual understanding of the material before introducing calculations. The tutorial curriculum materials consist of pre-tests, group worksheets, homework problems, and post-tests. These materials are scaffolded and prompt students to confront and resolve common misconceptions. Each tutorial section has one or two teaching assistants who are trained to guide students through the misconception confrontation process~\cite{Scherr2007EnablingMaterials}.

\subsubsection{ISLE}
The Investigative Science Learning Environment (ISLE) approach was developed at Rutgers University~\cite{Etkina2007InvestigativePhysics}. The ISLE approach is intended to be implemented in all parts of a course, but in some cases, it is possible to use the ISLE philosophy only in a lab. The ISLE approach helps students learn physics while treating them as novice scientists. Students are encouraged to use an iterative process in their learning, much like they would be expected to do in a scientific career. This process typically begins with observing a simple, carefully chosen ``observational experiment.'' Students, working in small groups, then try to explain the experiment based on their observations, and use their explanation to make predictions about the outcomes of new ``testing" experiments that they design. When there is a mismatch between the prediction and the outcome of the testing experiment, the students revise the explanation. Multiple explanations are encouraged for the observational experiments. To develop and test explanations, students use multiple representations. Unlike several of the other curricula studied in this project, ISLE focuses on building up a student's correct intuition rather than debunking misconceptions. It is used in large and small enrollment college physics courses and in many high school physics courses~\cite{Etkina2015MillikanPractices}.

\subsubsection{Modeling Instruction}
Modeling Instruction for university physics was developed collaboratively~\cite{PhysPortInstruction}, modeled after the high school Modeling Instruction curriculum developed by Wells and Hestenes~\cite{Wells1995AInstruction}.
Modeling Instruction is ideal for large, open classrooms with the ability for large group collaboration. It is a curriculum that focuses on having students build their fundamental understanding of physics from the ground up, by observing phenomena and then creating models to describe the phenomena. Students use multiple representations to explain their models, and deploy those models to future situations until they break. A typical day in class begins with a cooperative group activity. After the activity, students then engage in `whiteboard meetings', where they circle up with several other groups to share their results. This forces students to consolidate their ideas into an understandable, presentable format, and allows discussion between groups~\cite{Brewe2008}. There is very little lecture instruction; all material is learned through the activities and model building.

\subsubsection{Peer Instruction}
Peer Instruction was popularized by Eric Mazur at Harvard University~\cite{Mazur1997}. Peer Instruction is typically used in large lecture halls as a way to integrate active learning into traditional lecture-style courses, but can also be used in smaller courses. A typical cycle of instruction begins with approximately ten minutes of lecture followed by a clicker question. The question is posed, students answer individually, students discuss with their neighbors, and are sometimes allowed to answer again. This curriculum is commonly facilitated with personal-response systems, such as clickers, color-coded cards, or an online response program. While Peer Instruction is meant to refer to a specific routine and style of questioning~\cite{Crouch2007PeerOnce}, the implementation of Peer Instruction varies wildly between instructors~\cite{Turpen2009NotInstruction}.

\subsubsection{Context-Rich Problems}
Context-Rich Problems refers specifically to the Minnesota Model for Large Introductory Courses that was developed as a physics curriculum at the University of Minnesota~\cite{UniversityDevelopment}. Context-Rich Problems works within a standard course structure consisting of lectures, labs, and recitation/discussion sections. This pedagogy uses the cognitive apprenticeship model~\cite{CognitiveTeachers} with an emphasis on problem solving skills as a means to organize content in the course. For example, during lectures, the instructor solves problems using an expert-like framework, illuminating the hidden decision-making processes that are necessary in physics problems. During the labs and discussion sections, students practice solving problems while giving and receiving coaching from instructors and other students. This practice takes place in groups of 2 to 4, structured by the principles of cooperative group work ~\cite{CooperationPsycNET}. 

The context-rich problems that students encounter with this pedagogy differ from typical textbook problems in that they provide a realistic reason for calculating something. Students are encouraged to follow the same expert-like problem solving strategy their instructor demonstrates in lecture~\cite{Heller1992TeachingGroups}. These types of problems typically contain extraneous information, require the use of estimation, or require students to recall commonly known values. 
All of these activities immerse students in a ``culture of expert practice'' similar to a traditional apprenticeship.

\subsubsection{SCALE-UP}
Student Centered Activities for Large Enrollment Undergraduate Programs, or SCALE-UP, was developed by \citet{Beichner2007TheProject}. It is an integrated learning environment designed for large-enrollment physics classes with up to 100 students. The goal of SCALE-UP was to take a studio-style environment and make it accessible for larger courses. A typical SCALE-UP course does not include much lecture time, but instead relegates information transfer to assigned readings outside of class. This leaves class time for cooperative group problem solving, experiments, or answering questions that students may have. SCALE-UP refers more to the environment than the specific pedagogy; instructors are able to implement any pedagogy they wish, and have it easily translated to a large class via the room layout. A typical SCALE-UP classroom has large, round tables, capable of holding up to nine students. Within these tables, students are in teams of 3 with whom they solve problems and work on experiments. Having multiple teams at the same table allows for group-to-group interaction without wreaking havoc on classroom management. The room can either have whiteboards along the perimeter wall or individual whiteboards at the tables. These are used for group problem solving and sharing with the class. SCALE-UP has one session with the students, there is no separate lab or recitation section; everything is done in the same room when the lesson calls for it.

\section{Methodology}\label{sec:methods}

In this section, we describe how research sites were chosen and provide an overview of the selected site demographics. We broadly describe the social network and COPUS data collection protocols, then further describe the methods for each site. We also include descriptions of each pedagogical implementation.

\subsection{Research Site Selection}
To determine which curricula to study, we began by reviewing past New Faculty Workshops, hosted by the American Astronomical Society, the American Physical Society, and the American Association of Physics Teachers. There were three main criteria we looked for when deciding on pedagogies: Does the pedagogy have developed curriculum materials? Does the pedagogy have an established body of research? Is this pedagogy still widely used in actual physics classrooms? 

After identifying the six pedagogies used in this study, we reached out directly to the developers of the curriculum to identify high-fidelity implementations of each pedagogy. When possible, the institution that developed the pedagogy was used as a research site; however, this was not always feasible. Secondary institutions were identified via recommendation of instructors with extensive training or research experience with the pedagogy in question. At each study site, the hosting faculty member facilitated the class visits and was given a small stipend to compensate their time. Demographic information for the chosen sites can be seen in Table \ref{tab:Demographics}.

\begin{table*}[htbp]
  \caption{Institution-level demographics of research sites as provided by Integrated Post-Secondary Education Data System (IPEDS) \cite{U.S.DepartmentofEducationNationalCenterforEducationStatistics2018IntegratedIPEDS}. Class-level demographics were not available for this study. For demographics that do not sum to 100\%, the missing entries are `unknown.'  \label{tab:Demographics}}
  \begin{ruledtabular}
    \begin{tabular}{lllllll}
      & \textbf{Tutorials}
      & \textbf{ISLE}
      & \textbf{Peer Inst.}
      & \textbf{Context-Rich}
      & \textbf{Modeling}
      & \textbf{SCALE-UP}
      \\ 
      & 4-year & 4-year & 4-year & 2-year & 4-year & 4-year \\
      & Public & Public & Private & Public & Public & Public \\
      \hline
      Undergraduate Enrollment & 32,099 & 36,039 & 15,724 & 20,000 & 48,818 & 11,425 \\
      
      \hline
      Female & 53\% & 50\% & 48\% & 55\% & 57\% & 46\%\\
      Male & 47\% & 50\% & 52\% & 45\% & 43\% & 54\%\\
      
      \hline
      Full time & 92\% & 94\% & 89\% & 30\% & 57\% & 90\% \\
      Part time & 8\% & 6\% & 11\% & 70\% & 43\% & 10\% \\
      
      \hline
      American Indian or Alaskan Native & 0\% & 0\% & 0\% & 1\% & 0\% & 1\%\\
      Asian & 25\% & 27\% & 18\% & 5\% & 2\% & 1\%\\
      Black or African American & 3\% & 7\% & 7\% & 2\% & 12\% & 3\% \\
      Hispanic or Latino & 8\% & 13\% & 7\% & 36\% & 67\% & 3\% \\
      White & 38\% & 38\% & 52\% & 46\% & 8\% & 86\% \\
      Two or more races & 7\% & 4\% & 4\% & 5\% & 2\% & 3\% \\
      Race or ethnicity unknown & 3\% & 2\% & 3\% & 5\% & 1\% & 1\% \\
      Non-resident alien & 15\% & 9\% & 11\% & 1\% & 6\% & 2\%\\
      
      \hline
      24 and under & 92\% & 92\% & 86\% & 62\% & 76\% & 93\% \\
      25 and over & 8\% & 8\% & 14\% & 38\% & 24\% & 7\% \\
      
      \hline
      In state & 62\% & 83\% & 44\% & 93\% & 84\% & 35\% \\
      Out of state & 21\% & 7\% & 46\% & 0\% & 3\% & 64\% \\
      Foreign country & 16\% & 9\% & 9\% & 4\% & 12\% & 1\% \\
      
    \end{tabular}
  \end{ruledtabular}
\end{table*}

\subsection{Social Network Data Collection}
Research sites provided rosters of students enrolled in the introductory physics course for the term that would be observed. Students were invited via email to complete an online survey during the first week of the term and again at the end of the term. They were provided with a one-sentence introduction, informing them that we were studying how the social network of the classroom develops. Students were asked to select their name from a list of people enrolled in the class. Given the same list, they were presented with the question 
\begin{quote}
    ``Please choose from the list of people that are enrolled in your physics class the names of any other student with whom you had a meaningful interaction in class during the past week, even if you were not the main person speaking."
\end{quote}
Students self-identified what counted as a `meaningful' interaction. This approach to network data collection has been used successfully in physics education research previously~\cite{Zwolak2017,Dou2016BeyondLens, Brewe2010ChangingCommunities}. Instructors were asked to promote the survey in class to encourage participation. 

Students were included in the social network if they filled out the survey, or were named by someone who filled out the survey in either the pre or post distribution. If a student did not fill out the survey, or was not named by someone else in either the pre or post distribution, they were not included in the data. Additionally, students under 18 were not included as respondents due to Institutional Review Board (IRB) restrictions. Approximately 10-13\% of enrolled students marked the under 18 box in the pre or post survey distributions; however, these students can still be included in the network if someone else named them as connections. For example, as seen in Table \ref{tab:metrics}, SCALE-UP had an enrollment of 71 students and retained 69 nodes, despite having 7 students under 18 in the pre-survey and 11 students under 18 in the post-survey. The discrepancy between enrollment and number of nodes included in the network could have arisen from multiple factors; the student dropped the course before surveys were distributed, the student was under 18 and was not named by their peers, or the student did not fill out the survey and was not named by their peers. To be completely removed from the network requires that the student did not fill out the survey, was not named in the pre survey, and was not named in the post survey.

\subsection{COPUS Data Collection}
Each institution was visited for a full week during the term to conduct COPUS observations for as many sections as logistically possible. COPUS consists of 12 instructor and 13 student codes, for which the observer marks whether the coded behavior occurred or not during two-minute intervals. An activity is counted if the behavior occurs for at least five seconds during the two-minute interval. An example of a COPUS observation for a tutorial session can be seen in table \ref{tab:COPUSExample}.

\begin{table*}[th]
\caption{COPUS observation example. Time is measured in two-minute intervals.}
\label{tab:COPUSExample}
\begin{tabular}{c|ccccccccccccc|cccccccccccc}
 & \multicolumn{13}{c}{\textbf{Students Doing}} 
 & \multicolumn{12}{c}{\textbf{Instructor Doing}} \\
 \cline{2-26}
 \textbf{Time}& L & IND & CG & WG & OG & AnQ & SQ & WC & Prd & SP & T/Q & W & O & Lec & RtW & FUp & PQ & CQ & AnQ & MG & 1o1 & D/V & Adm & W & O \\
 \hline
 0-2   &&&& X &&&   &&&&&& X &&&&&&& X & X &&& X\\  \hline
 2-4   &&&& X &&& X &&&&&& X &&&&&&& X & X      \\  \hline
 4-6   &&&& X &&& X &&&&&& X &&&&&&& X & X      \\  \hline
 6-8   &&&& X &&&   &&&&&& X &&&&&&& X & X      \\  \hline
 8-10  &&&& X &&&   &&&&&& X &&&&&&& X & X      \\  \hline
 10-12 &&&& X &&& X &&&&&&   &&&&&&& X & X      \\  \hline
 12-14 &&&& X &&&   &&&&&&   &&&&&&&   & X      \\  \hline
 14-16 &&&& X &&& X &&&&&&   &&&&&&& X & X      \\  \hline
 16-18 &&&& X &&&   &&&&&&   &&&&&&& X & X &&& X\\  \hline
 18-20 &&&& X &&&   &&&&&&   &&&&&&& X & X      \\
 \hline
 \textbf{Total} & 0 & 0 & 0 & 10 & 0 & 0 & 4 & 0 & 0 & 0 & 0 & 0 & 5 & 0 & 0 & 0 & 0 & 0 & 0 & 9 & 10 & 0 & 0 & 2 & 0
\end{tabular}
\end{table*}

From these observations, we can compile the selected codes into COPUS profiles. The method of COPUS profile creation is not always clearly reported in literature that uses COPUS, making it difficult to compare analyses. It is typical of such studies to report percentages for each code, without a description of how ratios were taken or visualizations to infer the same information. As such, we explain our process here. For the CALEP project, we used the `bar chart' method, in which COPUS profiles were created by summing the number of marks in each column (indicated in the `total' row in table \ref{tab:COPUSExample}), and dividing by the number of intervals over which the observation occurred (10 intervals, using the same example). This method leads to the percentage of class time that a code was present. An example of a COPUS profile from CALEP can be seen in Table \ref{tab:COPUSProfileExample}. 

\begin{table}[th]
\caption{COPUS profile example. The total number of occurrences for each code was tallied in each column, and divided by the number of intervals. For this example, there were 10 intervals.\label{tab:COPUSProfileExample}}
\begin{ruledtabular}
\begin{tabular}{cc|cc}
 \multicolumn{2}{c}{Students Doing} 
 & \multicolumn{2}{c}{Instructor Doing} \\
 \hline
 L   & 0   & Lec & 0  \\
 IND & 0   & RtW & 0  \\
 CG  & 0   & FUp & 0  \\
 WG  & 1.0 & PQ  & 0  \\
 OG  & 0   & CQ  & 0  \\
 AnQ & 0   & AnQ & 0  \\
 SQ  & 0.4 & MG  & 0.9\\
 WC  & 0   & 1o1 & 1.0\\
 Prd & 0   & D/V & 0  \\
 SP  & 0   & Adm & 0  \\
 T/Q & 0   & W   & 0.2\\
 W   & 0   & O   & 0  \\
 O   & 0.5

\end{tabular}
\end{ruledtabular}
\end{table}

All observations were done in a live environment, in person, by the same observer. The official COPUS recording spreadsheet was used for these data collections~\cite{Smith2013}. A summary of the number of sections and the number of observations per section can be seen in Table \ref{tab:numbersections}. For the earliest observations (Tutorials and Peer Instruction), only the first 20 minutes were available for COPUS observations, but for every other section the entire class is recorded unless otherwise indicated. The 20 minute observations were still included in our analysis, as they were similar in structure to the entire session. As such, our implicit assumption based on observer and instructor feedback is that these provide an accurate representation of the class despite their shorter length. 

\begin{table}[h]
\caption{Number of course sections and observations for each pedagogy included in this study. We also include the approximate length, in minutes, of each observation.}\label{tab:numbersections}
\begin{ruledtabular}
\begin{tabular}{llcc}
 Pedagogy & Sections & Num.\ Obs. & Mins/Obs. \\
 \hline
  Tutorials & 1 Lecture & 1 & 20\\
  & 19 Tutorials & 1 & 20 \\
  \hline
  ISLE whole class  & 1 Lab & 1 & 176 \\
  & 4 Recitations & 1 & 80\\
  & 1 Lecture & 2 & 60\\
  \hline
  ISLE lab only & 5 Labs & 1 & 150--160\\
  \hline
  Modeling Instruction & 3 & 2 & 120--190\\
  \hline
  Peer Instruction   & 1     & 1   & 20 \\
  \hline
  Context-Rich & 2 Discussions & 1 & 50\\
  Problems & 4 Labs & 1 & 166 \\
  & 2 Lectures & 2 & 80\\
  \hline
  SCALE-UP & 1 & 3 & 76
\end{tabular}
\end{ruledtabular}
\end{table}


\subsection{Tutorials in Introductory Physics: Data Collection}
The tutorials site was a public institution in the Northwestern United States, classified as ``Doctoral Universities: Very High Research Activity" by the Carnegie Classification~\cite{Carnagie}. The course format included a lecture section, a laboratory section, and a tutorial section. The network surveys were distributed by lecture section, as all students in a lecture section were distributed into the same subset of tutorial sections. In the survey, names were grouped by tutorial section to facilitate students selecting peers they worked with in their tutorial class, but they were able to select anyone in the same lecture. Three lecture sections were included in this study, which correlated to 24 tutorial sections. The tutorial sections had approximately 20 students per section. Nineteen tutorial sections and one lecture section were observed with COPUS.

All tutorials occurred in the same classroom, and attendance was graded. Students were seated at small tables designed for four students, but group sizes ranged from two to five students. Students chose their own groups, and did not necessarily have to work with the same people each week. The tables had whiteboards in the middle to facilitate group discussion. Each section had one lead teaching assistant and one grader. The grader spent the first few minutes passing back graded papers, and then joined the lead TA as a teaching assistant. For the purposes of the COPUS observations, the lead TA was observed using instructor codes. Grader interactions were also coded, but not included in analysis. Tutorial sessions were 50 minutes long, but observation was only done for 20 minutes in the middle of the session.

\subsection{ISLE: Data Collection}
The ISLE site was a public institution in the Mid-Atlantic region of the United States, classified as ``Doctoral Universities: Very High Research Activity" by the Carnegie Classification~\cite{Carnagie}. There are two variations of ISLE at this institution, both of which were included in this study. The first is a lab-only implementation, where ISLE is used in the lab course and accompanied by a standard lecture and recitation. The second is a whole-course implementation, where ISLE principles are used across lab, recitation, and lecture.  

\subsubsection{Lab-Only Implementation}
In the lab-only ISLE implementation, the lab course is taken separately from a lecture and recitation course. The lab does not have to be taken in the same term as the lecture and recitation, so only the lab sections were observed with COPUS and given network surveys. Each section had approximately 28 students enrolled. Twenty lab sections were surveyed from this implementation. Of those, five sections were observed using COPUS.

The physics laboratory is a typical lab-bench set-up with one computer per station. Students work in groups of 2--3 to complete the activity, and use shared Google Docs to draft and submit their lab report. COPUS observations were taken for the entire 3 hour session. 

\subsubsection{Whole-Course Implementation}
The whole-class ISLE implementation consisted of a lecture component, and then small lab and recitation sections where group activities were performed. All sections of the whole-course implementation must be taken concurrently, so network surveys were distributed for each of the nine recitation sections as well as each of the nine lab sections. All students were in the same lecture, so a lecture survey was not distributed. Both lab and recitation sections had approximately 28 students per section. Two lecture sessions were observed with COPUS, as well as four recitations and one lab.  

The lecture was held in a stadium-seating auditorium style lecture hall. The lab and recitations were held in a separate lab/recitation room, which was set up with small round tables to encourage group cooperation. Each table held a group of 4 students, where they worked in pairs and shared with the other pair at the same table. On recitation days, the students worked through ISLE workbook activities. Recitations were an hour and a half long. On lab days, students performed experiments following the ISLE protocol. Labs were three hours long. COPUS observations were taken for the full length of each session.

\subsection{Modeling Instruction: Data Collection}
The Modeling Instruction site was a public institution in the Southeastern part of the United States, classified as ``Doctoral Universities: Very High Research Activity" by the Carnegie Classification~\cite{Carnagie}. The course used a studio format (integrated lab and lecture components) that met twice a week. Network surveys were distributed in each individual class section. Three sections were included in this study, with sizes ranging from 64 to 92 students. All three sections were observed using COPUS.

All course activity occurred in the same room, which was a large open format room with large tables for group activities. There was a large whiteboard and projectors at the front of the room. Students were distributed among the large tables in groups ranging from 2 to 6 people (assigned group size was 6 but due to absences could be as small as 2 on any given day). Student groups were assigned by the instructor and changed twice during the term. The tables had whiteboards in the middle to facilitate group discussion; in addition, the boards were used during `board meetings.' 
 
Within the groups, students ranged from working collaboratively on the assigned activity to working alone while sitting next to other students. However, the groups had to create a whiteboard summarizing their work to present to their peers during large whiteboard meetings, held 1--3 times during the class. Even if students worked alone on the activity, they were required to collaborate with their table-mates to create the whiteboard. During the whiteboard meeting, 3--5 groups made a large circle to discuss their work. Each group shared their whiteboards, and students discussed whether they obtained the same answers/results. Each section had several teaching assistants; in the COPUS observations, only the lead instructor was coded. Modeling Instruction sections varied from 2-3 hours in length depending on section, of which COPUS observations began 10--15 minutes into the session to allow students to get settled. The section presented in this manuscript was the shorter two-hour section.

Both days of instruction were aggregated into a single COPUS profile for each section. We felt it produced a more meaningful ``snapshot" of the curriculum to provide a week-long observation rather than a single class period.

\subsection{Peer Instruction: Data Collection}
The Peer Instruction site was a private institution in the Mid-Atlantic region of the United States, classified as ``Doctoral Universities: Very High Research Activity" by the Carnegie Classification~\cite{Carnagie}. The course format included a lecture, a laboratory section, and a recitation section. The network surveys were distributed by lecture section, since that is where the Peer Instruction curriculum was implemented. One section of Peer Instruction was surveyed and observed using COPUS, with an enrollment of 113 students.

Class was held in a large stadium seating lecture hall. Students sat wherever they wanted.  The laboratory and recitation sections were completely open during registration, so it was possible to have students mixing between lecture sections in the smaller lab and recitation sections. The survey rosters only included students in the lecture section in which they are enrolled in order to constrain the network. The network survey was distributed to the students in class as part of a ``Learning Catalytics" activity, which resulted in a significantly higher response rate. Lectures were 50 minutes long, of which twenty minutes were observed using COPUS, in the middle of the session.

\subsection{Context-Rich Problems: Data Collection}
The Context-Rich Problems site was a 2-year community college in the Mid-Pacific region of the United States, classified as ``Associate's Colleges: High Career \& Technical-High Traditional" by the Carnegie Classification~\cite{Carnagie}. The course format included a lecture, discussion section, and lab section. There were two distinct sections of students. Each section attended the same lecture and discussion, and then was split into two lab courses. Network surveys were distributed at the lecture/discussion level, with enrollments of 48 and 45 for each section. COPUS observations were performed for both discussion sections, all four lab sections, and two sessions of lecture per section, for a total of four lecture observations. 

The lab component was held in a traditional style lab room with lab benches that had computers at each station. Each group had 3--4 students. At the beginning of class, a question was projected on screen to help prime students for the activity. They were instructed to use whiteboards to design their experiment based on what they were trying to learn, and relate the data they were going to take to the concepts learned in class. Lab sections were 170 minutes long, of which the entire time was coded with COPUS.

The lecture was held in a small lecture hall with slightly tiered rows of stationary desks. The lecture was 80 minutes long, of which the entire time was coded with COPUS. The discussion section was held in the same room as the lecture. Despite the desks being immobile, students would physically turn in their seats to work in groups of 2--3 on a worksheet. They were provided whiteboards to facilitate discussion. An additional teaching assistant was present, although not included in the COPUS observations; they spent the entire time guiding discussions (MG and 1o1). Discussion sections were 50 minutes long, of which the entire time was coded. The students worked in the same groups in the lab and discussion sessions. These groups were assigned by the instructor and maintained for 3--4 weeks at a time before being reorganized so that students could work with different peers.

\subsection{SCALE-UP: Data Collection}
The SCALE-UP site was a public institution in the Great Plains region, classified as ``Doctoral Universities: High Research Activity" by the Carnegie Classification~\cite{Carnagie}. The course used a studio format  that met three times a week. There was only one section of the course, taught by one instructor, with an enrollment of 71 students.

The observed SCALE-UP curriculum was performed in a large room designed for active learning. There were several large, round tables with microphones at the center. The perimeter of the room was covered with white boards. There were television screens around the room in place of a projector, to allow for viewing of presentation materials from multiple angles. The instructor had a wireless microphone to allow for mobility without sacrificing sound quality.

Class was an hour and fifteen minutes long, of which the entire time was coded with COPUS. Three days of COPUS observations were aggregated into a single COPUS profile, as we felt it better represented the curriculum to give a week-long ``snapshot" instead of a single class period.

\section{Results}
While five of the six curricula had data collected from more than one section, we chose the section with the highest survey response rate for network analysis. Table \ref{tab:metrics} depicts the calculated network metrics for the social networks presented in this paper. These values will be discussed in detail in section \ref{sec:discussion-network}. Network diagrams for each pedagogy can be found in supplemental material. We give a brief discussion of notable features for each pedagogy, and use Modeling Instruction and SCALE-UP network diagrams to illustrate how the pedagogy can manifest in the network.

\begin{table*}[htbp]
\caption{Network metrics for six active learning curricula in physics: diameter, average degree, edge density, transitivity, size of giant component, number of nodes, and students enrolled in the section. The section with the highest response rate within each pedagogy is reported.}\label{tab:metrics}
\centering
\begin{ruledtabular}
\begin{tabular}{p{0.19\linewidth}p{0.08\linewidth}p{0.08\linewidth}p{0.08\linewidth}p{0.08\linewidth}p{0.1\linewidth}p{0.12\linewidth}p{0.08\linewidth}r}
Curriculum & Pre/Post & Diameter & Avg.\ Degree & Density & Transitivity & Giant Component & Number of Nodes & Enrolled \\ 
  \hline
Tutorials & Pre & 13 & 1.37 & 0.0094 & 0.134 & 41 & 147 & 171 \\

Tutorials & Post & 13 & 1.58 & 0.0108 & 0.264 & 67 & 147 & 171\\
\hline
ISLE lab only & Pre & 6 & 1.41 & 0.0541 & 0.231 & 15 &  27 & 28\\ 

ISLE lab only & Post & 7 & 1.93 & 0.0741 & 0.316 & 13 &  27 & 28 \\ 
\hline
ISLE whole class lab & Pre & 3 & 1.75 & 0.0761 & 0.414 & 5 &  24 & 28\\ 

ISLE whole class lab & Post & 4 & 2.25 & 0.0978 & 0.590 & 11 &  24 & 28 \\ 

ISLE whole class rec & Pre & 3 & 0.67 & 0.0333 & 0.000 & 4 &  21 & 28\\ 

ISLE whole class rec & Post & 4 & 1.43 & 0.0714 & 0.556 & 9 &  21 & 28\\ 
\hline
Modeling Instruction & Pre & 10 & 2.51 & 0.0339 & 0.179 & 62 &  75 & 77 \\ 

Modeling Instruction & Post & 5 & 5.60 & 0.0757 & 0.210 & 72 &  75 & 77 \\ 
\hline
Peer Instruction & Pre & 13 & 1.83 & 0.0174 & 0.233 & 61 & 106 & 116\\ 

Peer Instruction & Post & 12 & 2.23 & 0.0212 & 0.231 & 80 & 106 & 116 \\ 
\hline
Context-Rich Problems & Pre & 8 & 1.91 & 0.0434 & 0.198 & 28 &  45 & 48 \\ 

Context-Rich Problems & Post & 6 & 3.64 & 0.0828 & 0.245 & 41 &  45 & 48 \\ 
\hline
SCALE-UP & Pre & 7 & 1.71 & 0.0251 & 0.336 & 27 &  69 & 71 \\ 

SCALE-UP & Post & 8 & 4.09 & 0.0601 & 0.517 & 58 &  69 & 71  

\end{tabular}
\end{ruledtabular}
\end{table*}

\subsection{Network Results}
\subsubsection{Tutorials in Introductory Physics}

The network data for Tutorials in Introductory physics was collected at the lecture section level. The networks are characterized by groupings of 2--5 students, indicative of small group structure. There were also larger chain-like patterns, likely due to the geometry of the lecture hall and the interactivity of the lecture-based activities. Additionally, we saw linking between smaller clusters, indicating cross-group information transfer. The supplementary material has network diagrams colored by tutorial and lab section for the pre and post networks. We saw in both cases clustering by tutorial section, indicating that students largely worked with their tutorial group to learn physics, even in the lecture portion of the class. 

\subsubsection{ISLE: Lab-Only Implementation}
The early-term network has a large cluster of students, but the structure seemed to indicate that three students in particular were responsible for the majority of student interactions, as shown by their higher edge density. Meanwhile, in the post-network, the groupings of three students were much more pronounced, suggesting that lab groups were the dominant driver of student interactions.

\subsubsection{ISLE: Whole-Course Implementation}
{\em Recitation Network Data.}
The recitation sections for the COPUS and the network plots presented here are not from the same section. We chose to present the network with the highest response rate, which did not have a matching COPUS observation. Since all sections were taught by the same instructor and used the same activities, we assumed them to be similar enough to be considered side by side. The early term network shows a largely isolated student population. During recitation, students worked on workbook activities, so it could be explained as students working alone despite being at the same table as other students. In the end of term network, we can see some more prominent grouping, but still a large number of isolates. 

{\em Lab Network Data.}
The network plots for the lab sections show very distinct groupings, indicating that students did not speak to anyone outside of their assigned groups. As the term progressed, the distinct groups remained the dominant structure, but introduced some cross-group interaction.

\subsubsection{Modeling Instruction}
The early-term network plot in Fig.\ \ref{fig:MIPre} shows high levels of inter-connectivity. However, there were still two distinct ``islands", only connected to the main cluster via a single person. The late-term network shown in Fig.\ \ref{fig:MIPost} shows a much higher level of inter-connectivity; the density of this network more than doubled throughout the ten week period.

\subsubsection{Peer Instruction}
The early-term network had string-like structures in the network, likely caused by the geographical restriction that students were sitting in rows and thus limited to interactions with students in their immediate vicinity~\cite{Commeford2019PERC}\footnote{After publication, an error was discovered in the analysis. The original claim of zero triangles is incorrect, but the overall conclusion about structure of Peer Instruction being `string-like' remains unchanged. }. Branches could be explained by speaking to fellow students in front of or behind the student. There were a large percentage of isolated students, as students were not required to sit near or interact with their peers. The post-term network showed a very similar structure, but with slightly larger levels of connectivity, indicating that students were more likely to engage in discussion with their peers towards the end of the term.

\subsubsection{Context-Rich Problems}
The early-term network for one lecture/discussion section of the Context-Rich Problems curriculum had a couple of small groupings, likely from lab or discussion, while the rest of the class was lightly connected. The late-term network showed students who were much more heavily connected.

\subsubsection{SCALE-UP}

\begin{figure*}[htbp]

\begin{subfigure}[b]{0.45\linewidth}
\includegraphics[width=\linewidth]{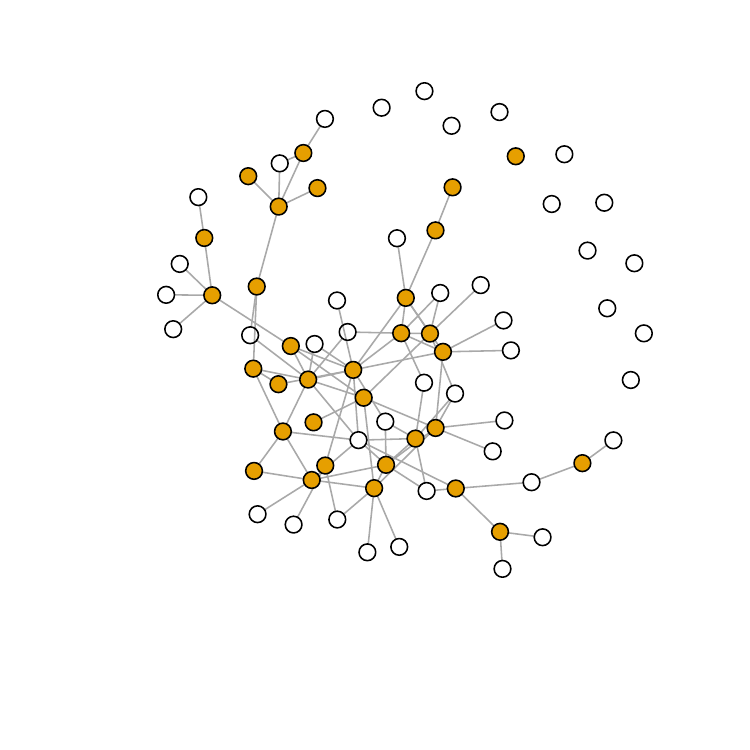} 
\caption{Modeling Instruction early-term network.}
\label{fig:MIPre}
\end{subfigure}
\begin{subfigure}[b]{0.45\linewidth}
\includegraphics[width=\linewidth]{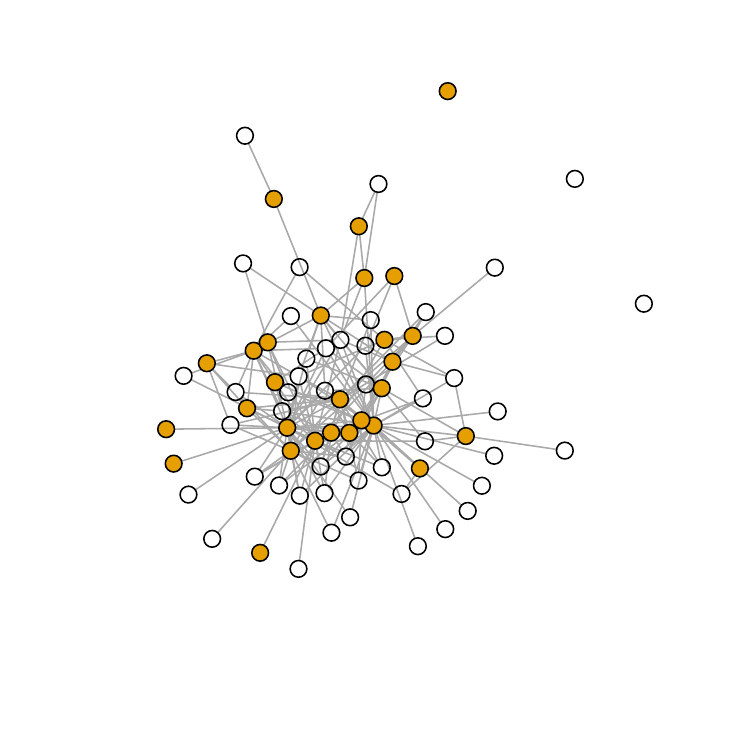}
\caption{Modeling Instruction late-term network.}
\label{fig:MIPost}
\end{subfigure}
\begin{subfigure}[b]{0.45\linewidth}
\includegraphics[width=\linewidth]{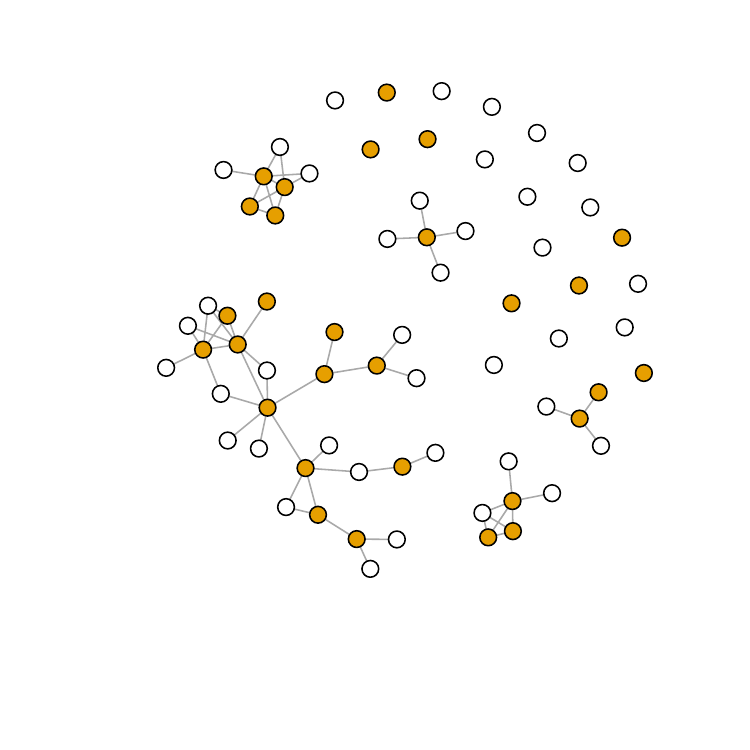} 
\caption{SCALE-UP early-term network.}
\label{fig:PreSCALEUPNetwork}
\end{subfigure}
\begin{subfigure}[b]{0.45\linewidth}
\includegraphics[width=\linewidth]{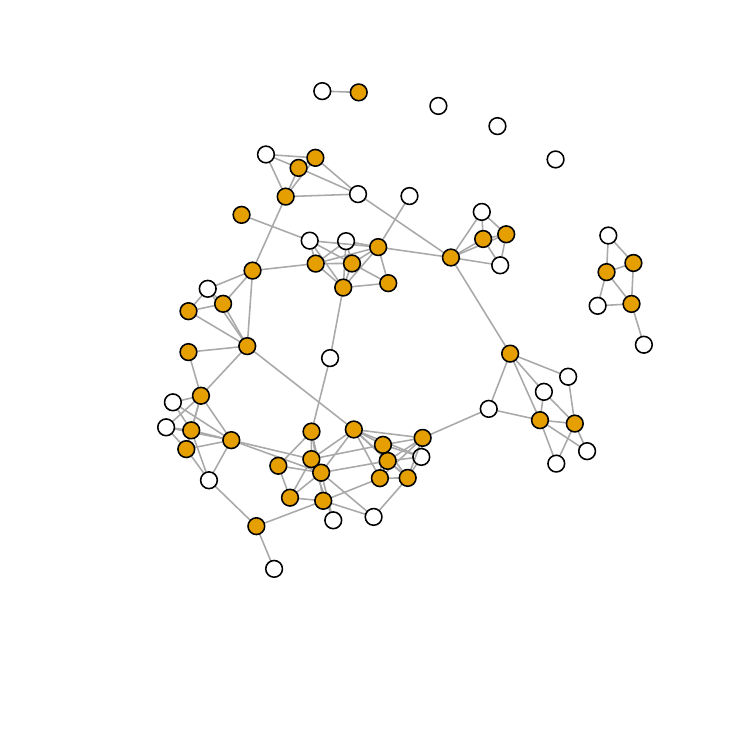}
\caption{SCALE-UP late-term network.}
\label{fig:PostSCALEUPNetwork}
\end{subfigure}

\caption{Early- and late-term networks for Modeling Instruction (top row) and SCALE-UP (bottom row). The two classes had similar sizes, but different network structures.}
\label{fig:image2}
\end{figure*}

The structure of the room is evident in the social network graphs, seen in figures \ref{fig:PreSCALEUPNetwork} and \ref{fig:PostSCALEUPNetwork}. The clustering of the students is indicative of the large table set-up of the classroom. 

Modeling Instruction and SCALE-UP are shown in figure \ref{fig:image2}. Despite both pedagogies being in large rooms with large groupings of students, they developed distinct network features indicative of the curriculum. Modeling Instruction focuses on whole-class discussions, and encourages cross-group interactions, while students in a SCALE-UP classroom largely remain in their groups. These features manifest in the networks---Modeling Instruction shows a tightly connected network, while SCALE-UP retains distinct clusters of students.
These differences are mirrored in the network measures on Table \ref{tab:metrics}: Late in the semester, Modeling Instruction has a lower diameter and a higher average degree, but the SCALE-UP network has a higher transitivity as its small-group clusters are more tightly connected.

\subsection{COPUS Results}

\subsubsection{Tutorials in Introductory Physics}
\begin{figure*}[h]
    \includegraphics[width=\linewidth,height=0.9\textheight,keepaspectratio]{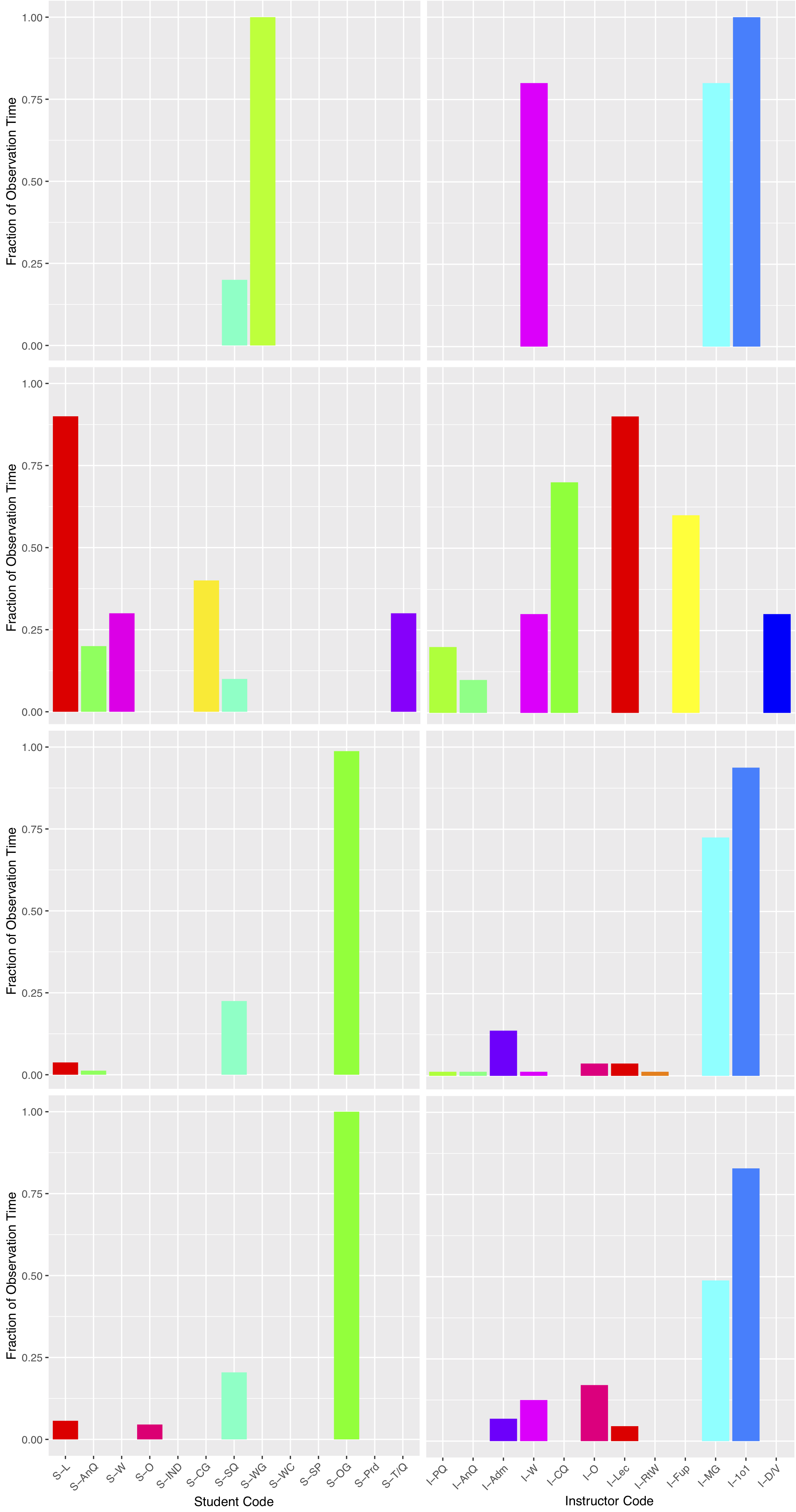}
    \caption{Top row: Tutorials in Introductory Physics, tutorial section. Second row: Tutorials in Introductory Physics, lecture section. Third row: ISLE, lab-only implementation. Last row: ISLE, whole class implementation, lab section.}
    \label{fig:mergedcopus1}
\end{figure*}

The tutorial section presented here was chosen at random from the subset of tutorial sections included in the top level lecture section. In the tutorial section, students worked in groups of 2-5 students. Within the groups, students ranged from working as a fully collaborative group on the tutorial worksheet assignment, to working alone while sitting next to other students. Figure \ref{fig:mergedcopus1} (first row, left panel) shows the student codes to illustrate the overwhelmingly common codes, SQ (student asks a question) and WG (working in groups). In these observations, the SQ COPUS code was only marked when a student explicitly raised their hand to ask the TA or grader a question, not questions during an already in progress one-on-one session. 

Figure \ref{fig:mergedcopus1} (first row, right panel) shows the instructor codes for this tutorial section to illustrate the three overwhelmingly common codes. The TA moved around the room and prompted students to work on the activity (MG) and stopped to have extended discussions with the student groups (1o1). The observations for the tutorial sections looked extremely similar, with small variations depending if the TA walked around the room and interjected themselves into student groups in an effort to drive conversation (coded as MG, moving around the room and guiding discussions, typically followed by 1o1 when a longer discussion arose), or if the TA waited at the front of the room until a student raised their hand with a question (coded as W-waiting). 

The tutorial section presented here was taken from the same lecture section as presented in the networks. However, the observed lecture was \emph{not} the same as presented in the network plots, but had the same instructor, so we assumed high levels of similarity between the two sections for COPUS. 
 
The observed lecture took place in a traditional, auditorium stadium-seating lecture hall with a large projector and chalk boards. The lecture began with a group activity about vectors, then used a response collection system to give students a short quiz. Students used phones or computers to answer the questions, and were allowed to work together as long as they gave their own answer. After the quiz, an interactive lecture was given using PowerPoint slides. The COPUS codes for the lecture section can be seen in the second row of figure \ref{fig:mergedcopus1}.

\subsubsection{ISLE: Lab-Only Implementation}

Figure \ref{fig:mergedcopus1} (third row, left panel) shows the student COPUS codes for one section of the ISLE lab-only implementation. The entire class period was spent working on the group lab activity (OG). Students would occasionally raise their hands to ask for help from the teaching assistant (SQ). In some sections, like the one shown here, the TA spent a few minutes during class going over concepts used in the lab activities, in which case the students listened (L) and answered questions posed to the entire class (AnQ).

Figure \ref{fig:mergedcopus1} (third row, right panel) shows the instructor COPUS codes from the observed TA behavior. The TA spent most of their time moving around the room and guiding the student activities (MG), and frequently stopped at groups to engage them in conversation about the activity (1on1). In some sections, like this one shown, the TA spent a few minutes going over concepts used in the lab activities, which was coded as Lec and RtW. During this short lecture time, the TA also posed questions to the entire class (PQ) and answered questions that the students had (AnQ). They also spent some time handing back papers and discussing grades with students (Adm), waiting for students to raise their hands (W), or talking to other TAs who happened to stop by (O).

\subsubsection{ISLE: Whole-Course Implementation}
\begin{figure*}[h]
    \includegraphics[width=\linewidth,height=0.9\textheight,keepaspectratio]{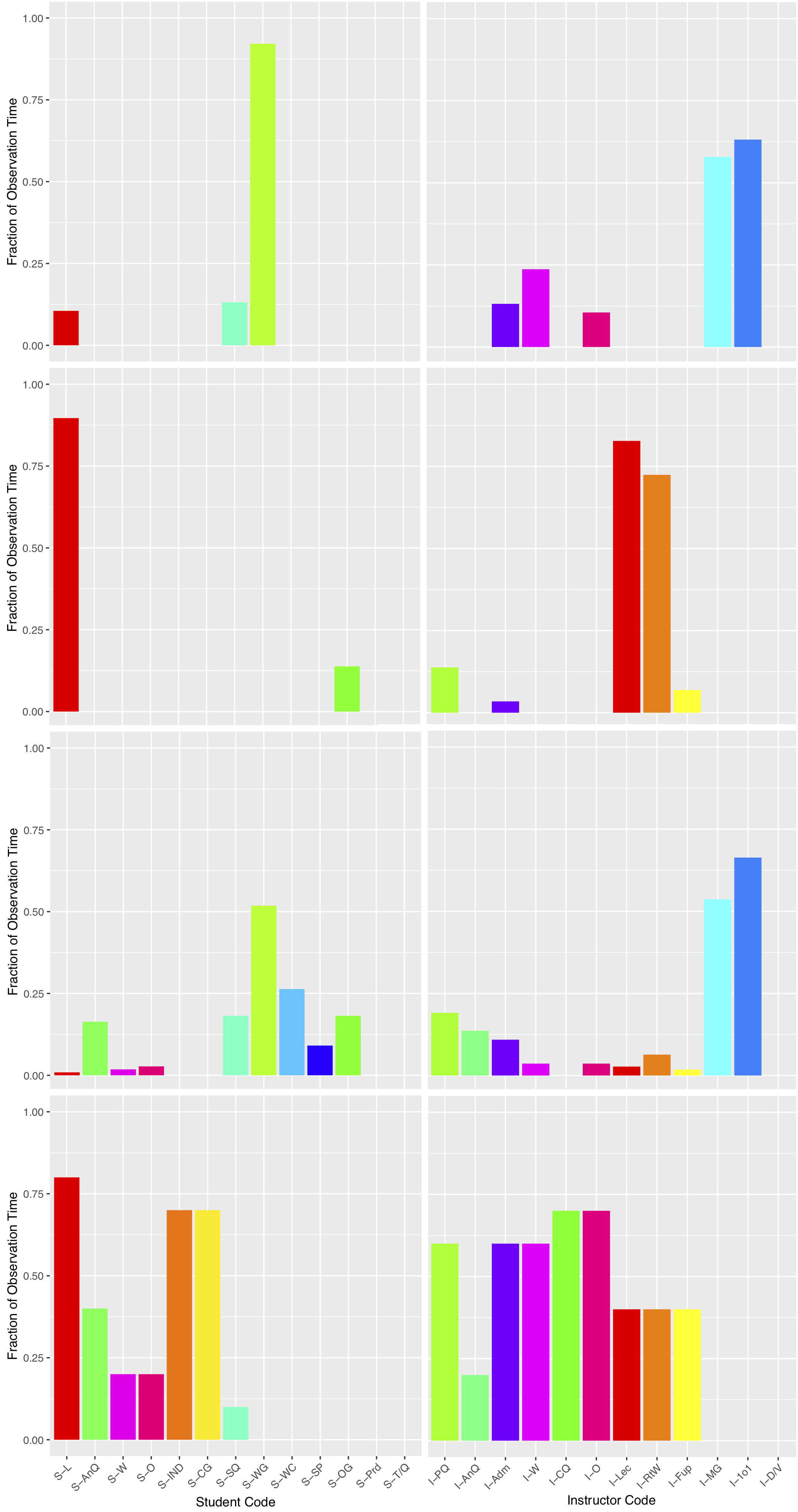}
    \caption{Top row: ISLE, whole class implementation, recitation section. Second row: ISLE, whole class implementation, lecture section. Third row: Modeling Instruction. Last row: Peer Instruction.}
    \label{fig:mergedcopus2}
\end{figure*}

{\em Recitation COPUS observations.}
Figure \ref{fig:mergedcopus2} (top row, left panel) shows the COPUS codes for the students during the recitation section of the whole-course ISLE implementation. The majority of the time was spent working on the active learning workbook activity (WG). Students occasionally raised their hands to ask the instructor a question (SQ), and spent some time listening to the instructor go over midterm instructions and instructions for registering for the next term (L).

Figure \ref{fig:mergedcopus2} (top row, right panel) shows the instructor COPUS codes for the recitation section. The instructor spent most of the time moving between groups and guiding the activity (MG), and frequently stopped for extended one on one discussions (1o1). Some time was spent describing how to register for the next term and midterm logistics (Adm). The remaining time was spent waiting for a student to raise their hand with a question (W) or talking to the TA (O).

{\em Lab COPUS Observations.}
Figure \ref{fig:mergedcopus1} (last row, left panel) shows the COPUS student codes for the whole-class implementation during the lab section. The majority of the lab course was spent working on the assigned activity (OG). Students occasionally raised their hands to ask the instructor a direct question (SQ). The instructor spent some time going over midterm instructions, with students listening (L). At the end of the day, students cleaned up their stations and prepared to go home (O). 
Figure \ref{fig:mergedcopus1} (last row, right panel) shows the instructor codes for the lab section of the whole-course ISLE implementation. The instructor spent the majority of the time moving around the room and guiding the activity (MG), and frequently stopped to have extended one-on-one discussions (1o1). Some time was spent describing the midterm (Adm), talking with the TA (O), briefly going over a common  misconception (Lec), or waiting at the side of the room for a group to raise their hands (W).

{\em Lecture COPUS Observations}
During the lecture, students listened to the instructor (L) and worked together on a short problem solving activity (OG). The instructor lectured (L) and wrote on the chalkboard (RtW), and posed an extended question to the entire class (PQ). Students were not expected to respond with clickers, but instead put their answer on a paper to be handed in. The rest of the time was spent following up the extended question (FUp) and talking about the midterm (Adm). Network data was not collected at the lecture level. The COPUS profile for the lecture portion of this implementation can be seen in figure \ref{fig:mergedcopus2} (second row).

\subsubsection{Modeling Instruction}
We present the section with the highest response rate on the network survey. Two COPUS observations were done for each section. The two corresponding COPUS observations were aggregated into one graph to represent a week of class time. This particular section had 11 groups of students, ranging in size from 2 to 6 students. There were three teaching assistants. The first day of class was spent working on a worksheet in the small groups, and the second day extended that activity to include an investigative lab experiment. 

Figure \ref{fig:mergedcopus2} (third row, left panel) shows the student COPUS codes for the Modeling Instruction course. While students spent most of the time working in groups on their worksheets (WG), they also spent time having whole class discussions via whiteboard meetings (WC). The experiment was coded as OG for other group activity. SQ was only coded when a student explicitly raised their hand to draw a TA to the group. 

Figure \ref{fig:mergedcopus2} (third row, right panel) shows the instructor COPUS codes. Most of the time was spent moving around the room, guiding the activities (MG) and frequently stopping for extended one on one discussions (1o1). During the whiteboard meetings, the instructor would frequently lead the discussion by posing questions (PQ) to the students or answering questions the students still had after the discussion.

\subsubsection{Peer Instruction}
The Peer Instruction COPUS data had the greatest range of significantly present activities recorded. Figure \ref{fig:mergedcopus2} (last row, left panel) shows the COPUS student codes for the observed Peer Instruction section. Students spent most of the class listening and taking notes (L), or responding to clicker questions. Clicker questions were evenly split between individual thinking (Ind) and group discussion (CG), indicating that the instructor followed the suggested Peer Instruction model for clicker questions, which included time for both individual and group discussion during the answer portion. There were also a few instances of students asking questions (SQ) and students being called on to answer questions (AnQ). The rest of the time was spent waiting for the instructor to enable the clicker response system (W).

The instructor COPUS codes can be seen in figure \ref{fig:mergedcopus2} (last row, right panel). The instructor spread their time across multiple activities. Less than half of the class time was spent lecturing and writing on the board (Lec and RtW). More than half of the time was spent posing individual questions (PQ) or clicker questions (CQ). After a question was posed, the instructor spent time following up by going over the answer (FUp) and answering questions the students still had (AnQ). For this class period, a lot of time was spent waiting for the students to answer the questions (W) or working through technical issues and activating the student response system (O). Administrative tasks were also performed (Adm) in the form of handing back midterm exams while the students were thinking through clicker questions.

\subsubsection{Context-Rich Problems}
\begin{figure*}[h]
    \includegraphics[width=\textwidth,height=0.9\textheight,keepaspectratio]{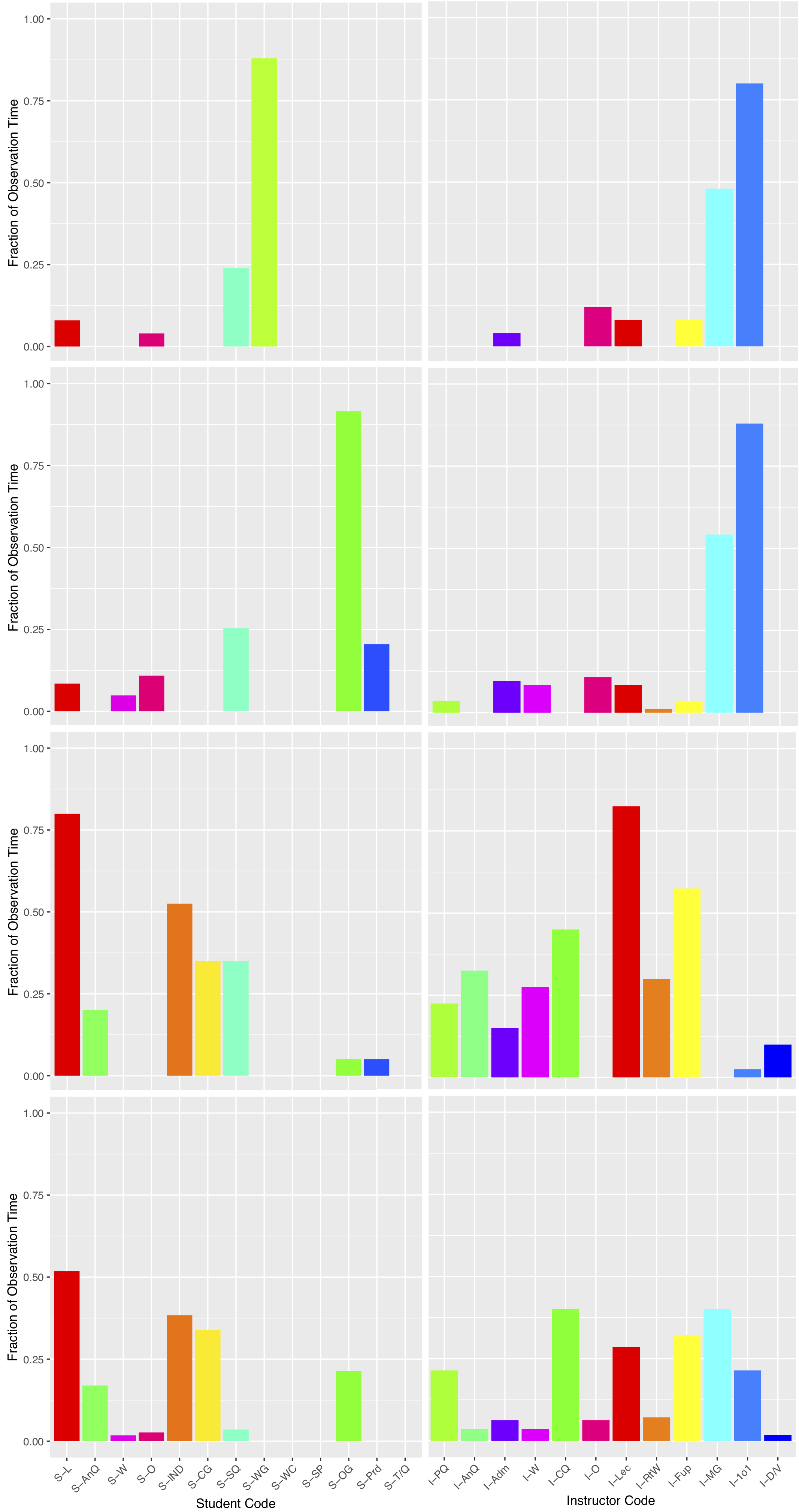}
    \caption{Top row: Context-Rich Problems, discussion section. Second row: Context-Rich Problems, lab section. Third row: Context-Rich Problems, lecture section. Last row: SCALE-UP.}
    \label{fig:mergedcopus3}
\end{figure*}

COPUS observations were taken for each part of the Context-Rich Problems curriculum. The discussion section observations can be seen in Figure \ref{fig:mergedcopus3} (top row). Students spent the majority of the time working in groups on the worksheet activity (WG). They frequently raised their hand to ask the TA/instructor a question (SQ). The instructor occasionally made announcements about the activity to the whole class, which students listened to (L). Whiteboards were available to facilitate group cooperation, which were gathered and put away at the beginning and end of class (O). The instructor spent the majority of the time walking around the room and guiding the activity (MG), frequently stopping to have extended discussions with student groups (1o1). The instructor also handed back papers (Adm) and went over points of difficulty with the activity to the whole class (Lec and FUp).

The lab section observations can be seen in Figure \ref{fig:mergedcopus3} (second row). Students spent the first portion of the class listening to the instructions (L). Students then solved a context-rich problem in their groups that was related to the lab activity (OG), and used the problem to predict the outcome of their experiment (Prd). The lab activity was then performed for the rest of the period (OG). When students had questions, they raised their hands to attract the instructor to their group (SQ). There were limited numbers of some apparatus, so groups would have to wait for them to become available (W). At the end of the class, the activity was cleaned up for the next class (O).

The instructor began the class by giving instructions for the activity and reviewing key concepts that would be used (Lec, RtW). They then posed a context-rich problem to the students that was related to the activity and discussed the experiment as a class (PQ, FUp). Once the activity began, the instructor moved around the room (MG) and frequently stopped for extended small group discussions (1o1). Equipment sometimes needed troubleshooting (O), and papers were passed back to the students (Adm).

The lecture section observations can be seen in Figure \ref{fig:mergedcopus3} (third row). The students listened to the instructor (L), answered directed questions (AnQ), and answered clicker questions individually and as pairs (CG, Ind). Students also asked the instructor questions with the whole class listening (SQ). When demos were presented, students were asked to predict the outcome (Prd) and solve a related context-rich problem in pairs (OG). 

The instructor would lecture and write on the board (Lec, RtW) to present new content, and to follow up questions (FUp). Numerous clicker questions were posed (CQ), as well as non-clicker questions (PQ). The instructor would use demonstrations to illustrate non-clicker Context-Rich Problems (D/V), and answered individual student questions (AnQ). Papers were also handed back during group thinking time (Adm).

\subsubsection{SCALE-UP}

The week-long observation period allowed for three class sessions to be documented. The COPUS data shown in figure \ref{fig:mergedcopus3} (last row) was aggregated over all three observation periods to represent a week's worth of class time.

Similar to other curricula already discussed, students solved problems in groups using a whiteboard (OG). However, this whiteboard was on the perimeter of the room, so students physically got up and walked to the whiteboards. Students also spent time listening to short lectures (L), answering clicker questions both alone and in their groups (Ind and CG), and asking questions to the instructor (SQ). When the student groups were called on during a whiteboard or clicker activity, it was marked as AnQ. 

The instructor codes hit all of the categories available with COPUS. There was some time spent giving short lectures via powerpoint (Lec), posing individual (PQ) and clicker questions (CQ), and following up those questions (FUp) with discussion and sometimes whiteboard explanations (RtW). During problem solving or clicker question time, the instructor moved around the room (MG) and engaged in discussions with the individual groups (1o1). During part of the lecture period, a short PhET~\cite{Perkins2006PhET:Physics} simulation was shown (D/V).


\section{Discussion}\label{sec:discussion}

The goal of this project was to develop a vocabulary to describe active learning pedagogies as individual entities. In this section, we will discuss how the network metrics varied with curricula and noticeable trends. We also discuss the overall trends within the COPUS profiles.

\subsection{Network Analysis}\label{sec:discussion-network}
Table \ref{tab:metrics} shows the calculated network metrics for the presented curricula. 

Diameter is tied to class size, with larger networks tending to have larger diameter, so the absolute value 
should not be compared across curricula. However, the change in diameter from the beginning to the end of term for each network varied with curriculum. Most curricula had a change in diameter of $\pm$ 1-2 students. This suggests that the overall style of interaction between the students remained largely unchanged from the early to late-term networks. However, we saw a large decrease in diameter with Modeling Instruction, indicative of a significantly more tight-knit class community at the end of the term. Modeling promotes not only in-group cooperation, but it also brings more students together to discuss concepts via whiteboard meetings. Thus, a large change in diameter could be a distinguishing feature of a Modeling Instruction classroom.

Average degree, on the other hand, is not as limited by class size. The average degree increased throughout the term in all curricula. This makes sense qualitatively; as students became more familiar with their peers, they formed more connections, regardless of curriculum choice. We do notice, however, that the largest gains in average degree occurred with Modeling Instruction (+3.1), SCALE-UP (+2.4), and Context-Rich Problems (+1.7). The remaining curricula had gains ranging from 0.21-0.76, averaging less than one new connection per person throughout the term. This result suggests that Modeling, SCALE-UP, and Context-Rich Problems are curricula that foster building new connections while retaining old connections made previously in the term.  

With an increase in average degree, the density of connections increased in all pedagogies as well. However, since density is also directly correlated to class size (larger networks tend to have lower density), we cannot make conjectures about pedagogical influence on this metric.

Transitivity, which can be loosely described as the base-level of collaboration between three students, increased for all curricula except Peer Instruction, which remained approximately constant. All curricula have a focus on collaborative group interactions, with the exception of Peer Instruction, which instead promotes partner interactions. The spatial limitations of the Peer Instruction classroom may have also inhibited transitivity growth, as the ability for students to engage with more peers than are present in their immediate vicinity was hindered. As such, a stable level of transitivity throughout the term may be indicative of a Peer Instruction classroom. 

The giant component in all but ISLE lab-only networks increased at the end of the term, indicative of more connectivity throughout the entire class. As with diameter, the absolute value of the giant component depends on class size, so change is more meaningful when discussing this metric. 

While we did notice promising features during this descriptive analysis, it is unclear whether certain features arise due to the curriculum itself or the classroom layout. If we compare the post-network diagrams from Modeling Instruction (Fig. \ref{fig:MIPost}) and SCALE-UP (Fig. \ref{fig:PostSCALEUPNetwork}), we see completely different structures despite similar classroom layout. Both Modeling Instruction and SCALE-UP had students sitting at large tables and working collaboratively. However, the Modeling Instruction social network developed into a tightly knit learning community, while the SCALE-UP social network retained distinct groupings based on the students' physical locations in the room. This may be attributed to the addition of the whole-class whiteboard meetings in Modeling, whereas SCALE-UP retains individual group identities during discussions.


While this project has showed promise for a method of describing active learning curricula independent of lecture methods, there are some serious limitations that need to be addressed. For one, the need for students to respond to a survey to develop social networks introduces opportunity for data loss via poor response rates. In a few cases, the instructor provided time in class to fill out the survey. Response rates for sections that were allotted class time to fill out the survey were significantly higher than those that did not. While network metrics are typically robust to missing data~\cite{Smith2013StructuralRandom, Smith2017NetworkMeasurement} due to the reciprocity of ties between students, poor response rate can render an incomplete picture. 

We combated low response rates in a few ways. Reciprocity of ties allowed us to include members of the class who were underage or declined to participate in the survey. This means that if a student did not fill out the survey, or were later cut for being under 18, they could still be presented in the network if someone else named them as a meaningful interaction. Second, we used an undirected network, meaning that if person A named person B, the edge between person A and person B existed, even if person B did not also name person A. Finally, data cuts only included students that filled out either the pre or post-survey, or were named in the pre or post-survey. 

\subsection{COPUS}
At this level of descriptive analysis, we see that the student COPUS profiles in Tutorials, ISLE recitation, and Context-Rich Problems discussion are very similar. COPUS only has two codes to refer to student group activities: working in groups on a worksheet (WG) and other group activity (OG). The code WG was appropriate for Tutorials, ISLE recitation, and Context-Rich Problems discussion, rendering similar profiles. However, all other group activities were indistinguishable within the OG category. `Other group activity' was coded for experiments, lab report creation, white board collaboration, and non-clicker problem group discussion. While two codes for small-group student collaboration may be appropriate for a Peer Instruction setting, which COPUS was designed for, we lost the capability to meaningfully distinguish curricula at the student-group level.

Similarly, the COPUS code O (Other) was marked for numerous student and instructor activities. For the instructor, this code included instances of bathroom breaks, conversations with the TA or another instructor, organizing lab materials, eating a snack during an extended teaching block, and troubleshooting equipment. For students, this code was used for clean-up, gathering materials, miming unit-vectors, turning in homework, or taking short breaks. The wide range of activities that can be classified as `O' is troublesome, as instructional activities can sometimes fall into this category that is largely dominated by non-instruction behaviors.

\section{Conclusion}
Despite losing resolution at the student-group level, and the broad brush of activities included as `other', it is still possible that COPUS will be able to differentiate between curricula as a whole; further study using latent profile analysis is underway~\cite{commeford2021characterizing}. However, while a valuable tool for interactive lecture environments, COPUS failed to distinguish student group activities. This could suggest that these pedagogies are not as distinct as we like to think, or that COPUS is not detailed enough for making these measurements.

Network analysis illuminated possible distinguishing features, such as a large decrease in network diameter with Modeling Instruction, and static transitivity with Peer Instruction. Now that we see promise in using this method to characterize active learning environments, a larger scale study would be advised. Additionally, we plan to use exponential random graph modeling to determine if these metric trends are coincidental or a feature of the pedagogy~\cite{Robins2007}. 

Our goal with this project was to develop a vocabulary to discuss active learning curricula independently of lecture, which we have begun by using COPUS observations and network analysis. More in-depth analysis of both network metrics and COPUS profiles will be featured in future papers.

\section{Acknowledgements} We would like to thank the institutions that participated in data collection and observation arrangements. We would also like to express our heart-felt thanks to the instructors that allowed us to infiltrate their classrooms for a week, with an additional thank you to those who also provided thorough and thoughtful feedback on this paper. Lastly, we would like to thank the National Science Foundation (DUE 1711017 and 1712341) for their generous support of this project.

\bibliography{main} 

\end{document}


\title{Supplemental Material}
\maketitle

\begin{figure*}[htbp]

\begin{subfigure}[b]{0.45\linewidth}
\includegraphics[width=\linewidth]{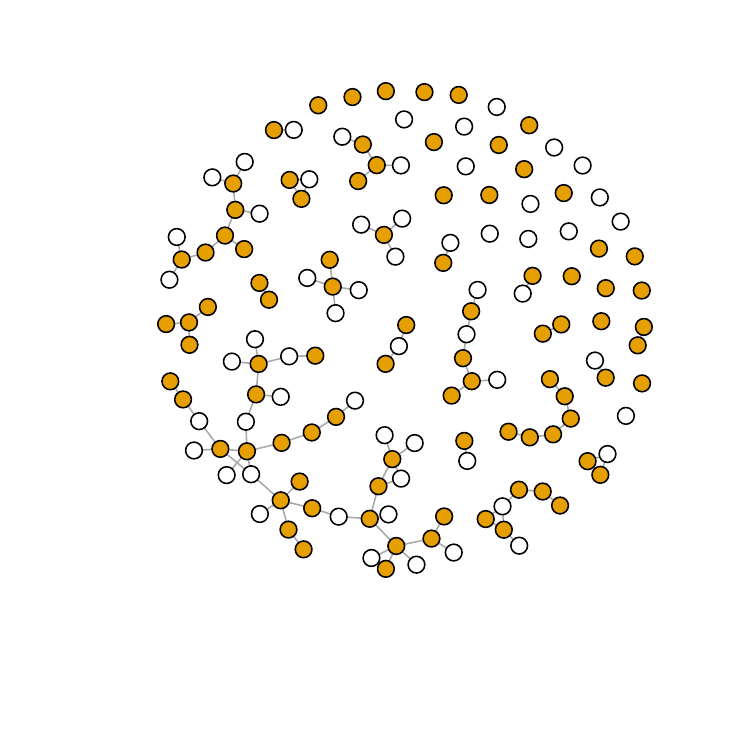} 
\caption{Beginning of term social network for Tutorials in Introductory Physics. }
\label{fig:PreTutorialNetwork}
\end{subfigure}
\begin{subfigure}[b]{0.45\linewidth}
\includegraphics[width=\linewidth]{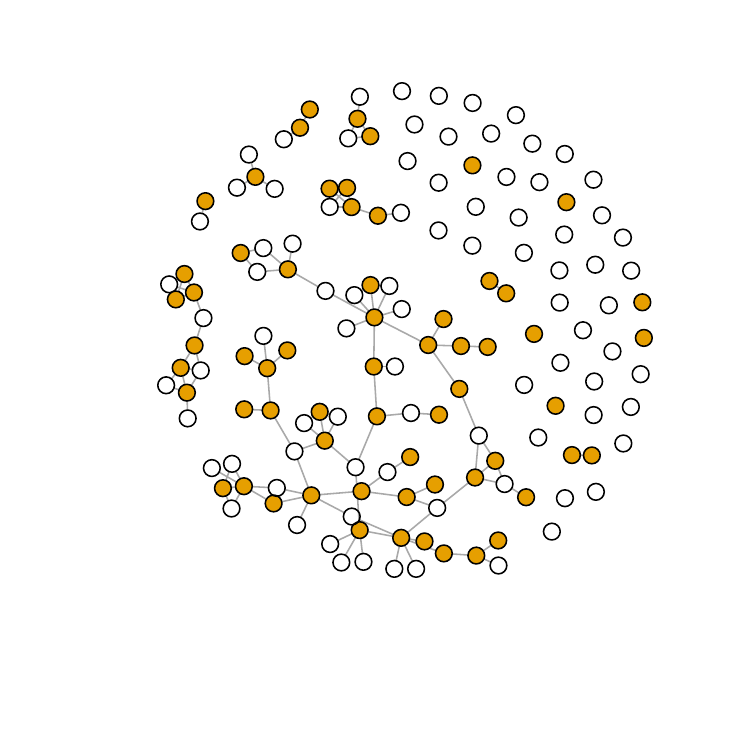}
\caption{End of term social network for Tutorials in Introductory Physics. }
\label{fig:PostTutorialNetwork}
\end{subfigure}

\caption{Pre and post-term social networks for Tutorial in Introductory Physics. The network survey was distributed at the lecture level. Small groupings of students are visible, likely due to tutorial groups. The color shading on these plots indicates which students filled out the survey.}
\label{fig:Networks-tutorials}
\end{figure*}

\begin{figure*}[htbp]
\begin{subfigure}[b]{0.45\linewidth}
    \includegraphics[width=\linewidth]{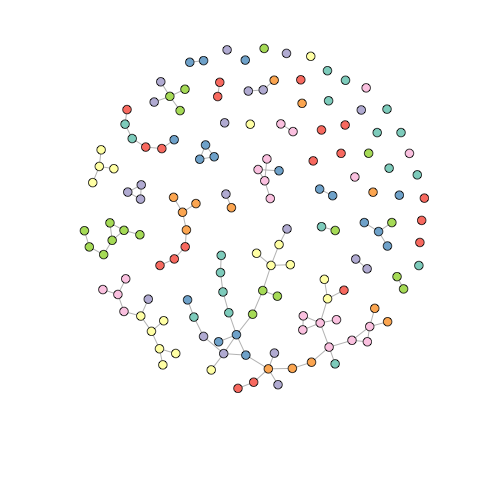}
    \caption{Beginning of term social network for Tutorials in Introductory Physics, colored by tutorial section. }
    \label{fig:TutorialsectioncolorNetworkpre}
\end{subfigure}
\begin{subfigure}[b]{0.45\linewidth}
    \includegraphics[width=\linewidth]{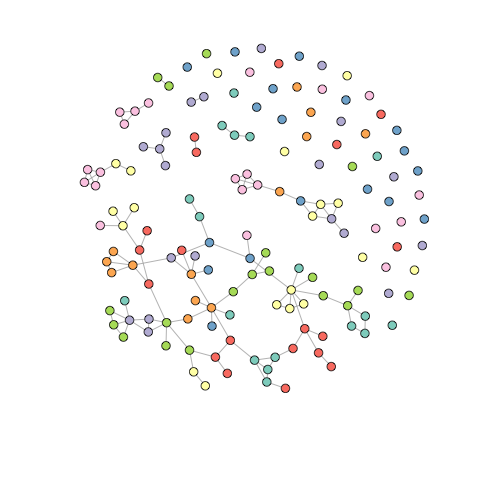}
    \caption{End of term social network for Tutorials in Introductory Physics, colored by tutorial section.}
    \label{fig:TutorialsectioncolorNetworkpost}
\end{subfigure}
\caption{Pre and post-term social networks for Tutorials in Introductory Physics, colored by tutorial section. We see clustering by color, indicating that students form connections with peers in the same tutorial section.}
\end{figure*}

\begin{figure*}[htbp]
\begin{subfigure}[b]{0.45\linewidth}
    \includegraphics[width=\linewidth]{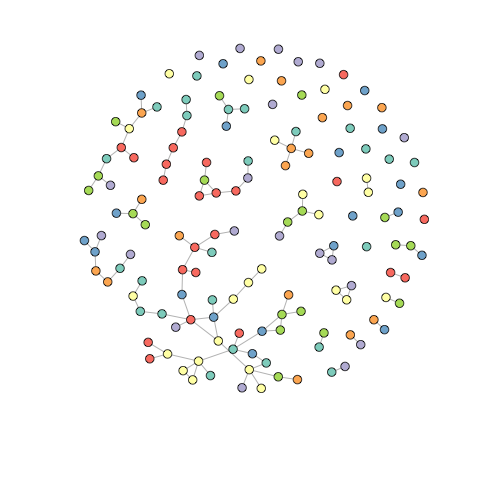}
    \caption{Beginning of term social network for Tutorials in Introductory Physics, colored by lab section.}
    \label{fig:labsectioncolorNetworkpre}
\end{subfigure}
\begin{subfigure}[b]{0.45\linewidth}
    \includegraphics[width=\linewidth]{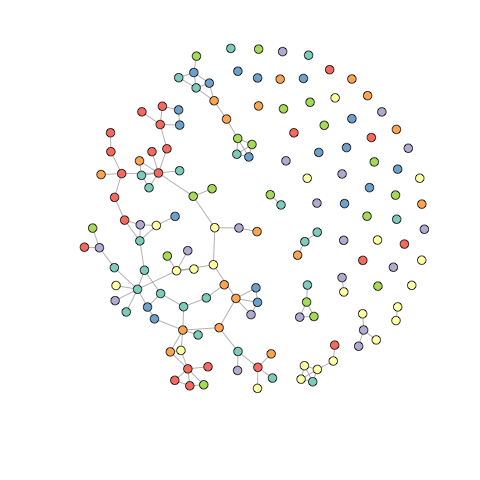}
    \caption{End of term social network for Tutorials in Introductory Physics, colored by lab section.}
    \label{fig:labsectioncolorNetworkpost}
\end{subfigure}
\caption{Pre and post-term social networks for Tutorial in Introductory Physics, colored by lab section. While there is still some clustering by color, the tutorial section seems to be the main driver for student connections.}
\end{figure*}

\begin{figure*}[htbp]

\begin{subfigure}[b]{0.45\linewidth}
\includegraphics[width=\linewidth]{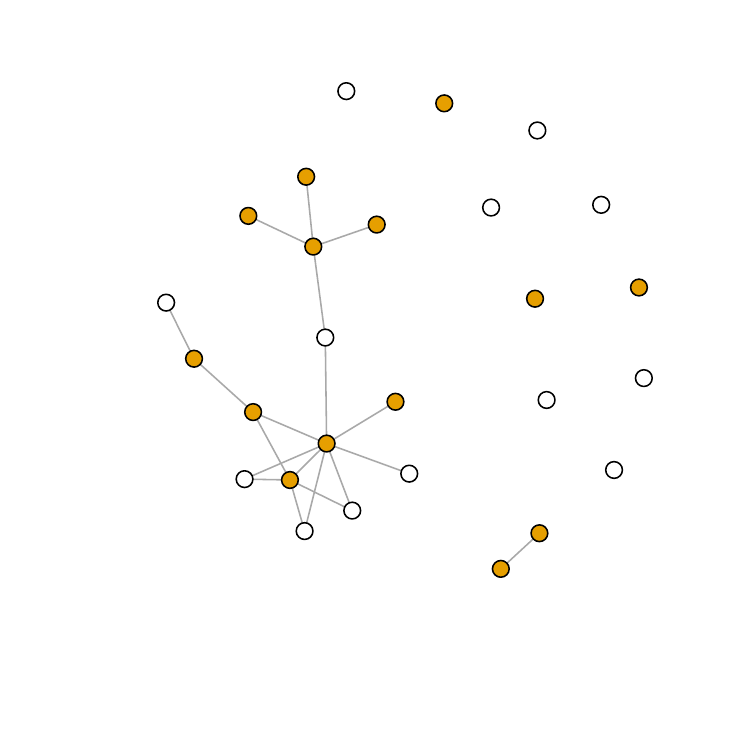} 
\caption{Beginning of term social network for one lab section of the lab-only ISLE curriculum.}
\label{fig:PreISLELabOnlyNetwork}
\end{subfigure}
\begin{subfigure}[b]{0.45\linewidth}
\includegraphics[width=\linewidth]{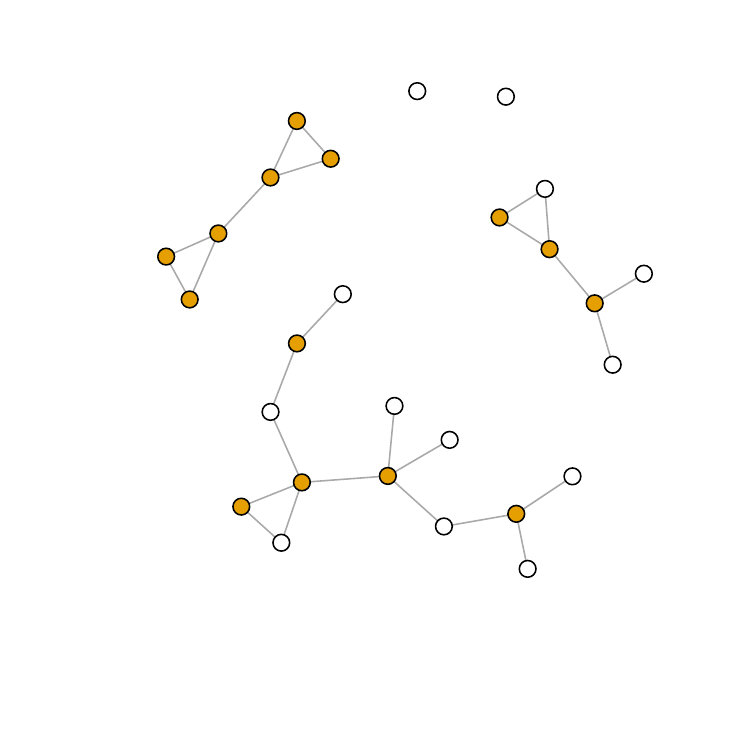}
\caption{End of term social network for one lab section of the lab-only ISLE curriculum.}
\label{fig:PostISLELabOnlyNetwork}
\end{subfigure}

\caption{Pre and post-term social networks for the lab-only implementation of ISLE. The network survey was distributed within individual lab sections. The color shading on these plots indicates which students filled out the survey. }
\label{fig:Networks-ISLElabonly}
\end{figure*}

\begin{figure*}[htbp]

\begin{subfigure}[b]{0.45\linewidth}
\includegraphics[width=\linewidth]{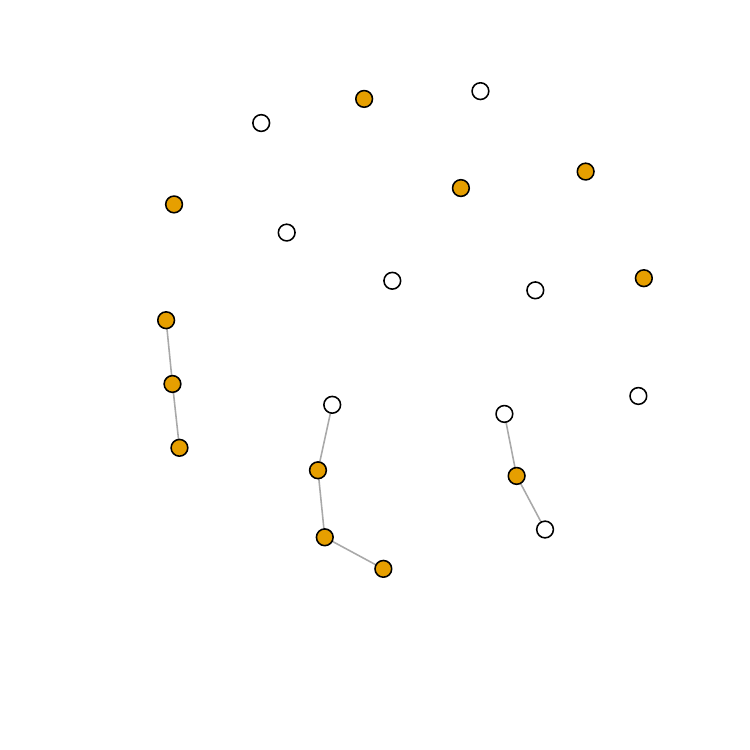} 
\caption{Beginning of term social network for one recitation section of the whole-course ISLE curriculum.}
\label{fig:PreISLERecitationNetwork}
\end{subfigure}
\begin{subfigure}[b]{0.45\linewidth}
\includegraphics[width=\linewidth]{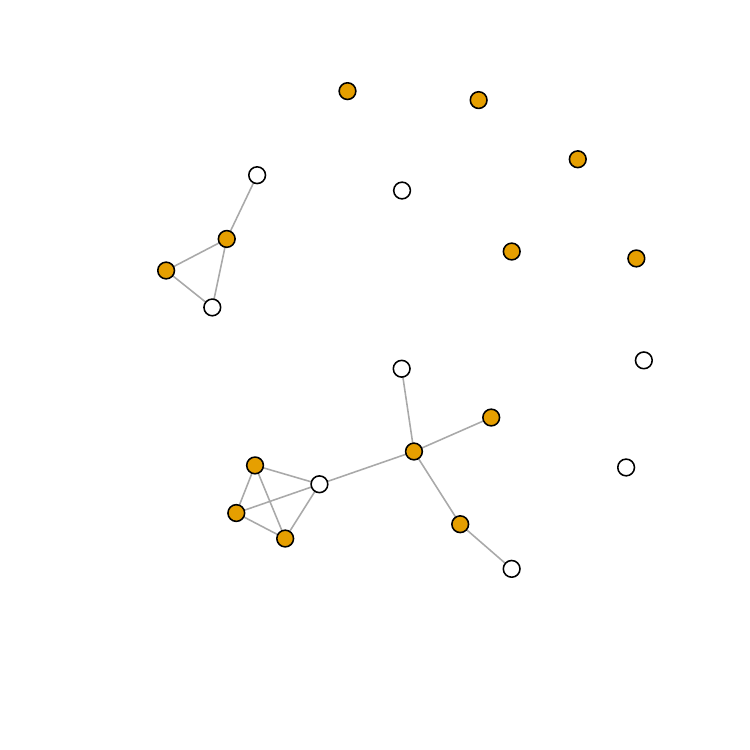}
\caption{End of term social network for one recitation section of the whole-course ISLE curriculum.}
\label{fig:PostISLERecitationNetwork}
\end{subfigure}

\caption{Pre and post-term social networks for a recitation section of the whole-course implementation of ISLE. The network survey was distributed in the individual recitation sections. Small pairings of students are visible, likely due to partner collaboration on recitation activities. The color shading on these plots indicates which students filled out the survey.}
\label{fig:Networks-ISLEwholecourseRec}
\end{figure*}

\begin{figure*}[htbp]

\begin{subfigure}[b]{0.45\linewidth}
\includegraphics[width=\linewidth]{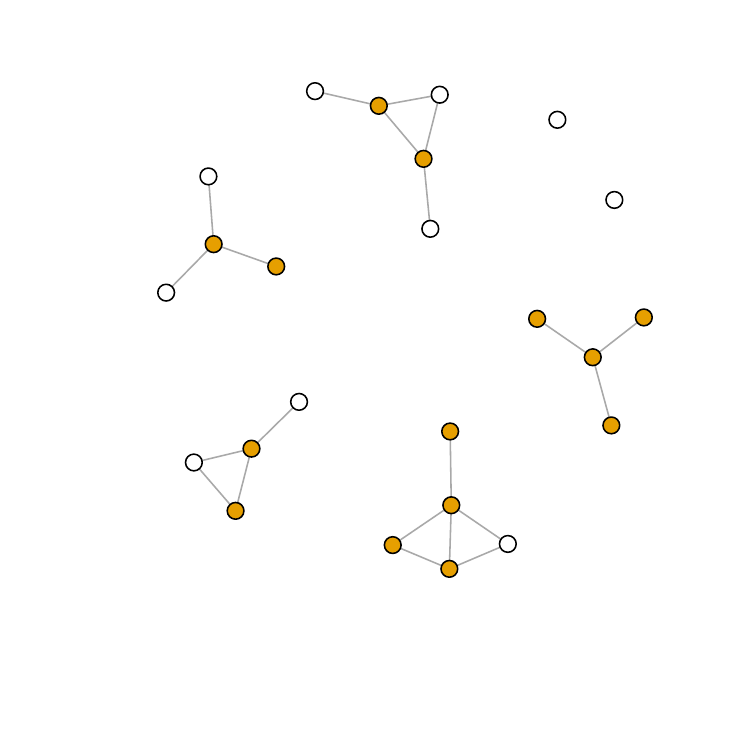} 
\caption{Beginning of term social network for one lab section of the whole-course ISLE curriculum.}
\label{fig:PreISLEFullLabNetwork}
\end{subfigure}
\begin{subfigure}[b]{0.45\linewidth}
\includegraphics[width=\linewidth]{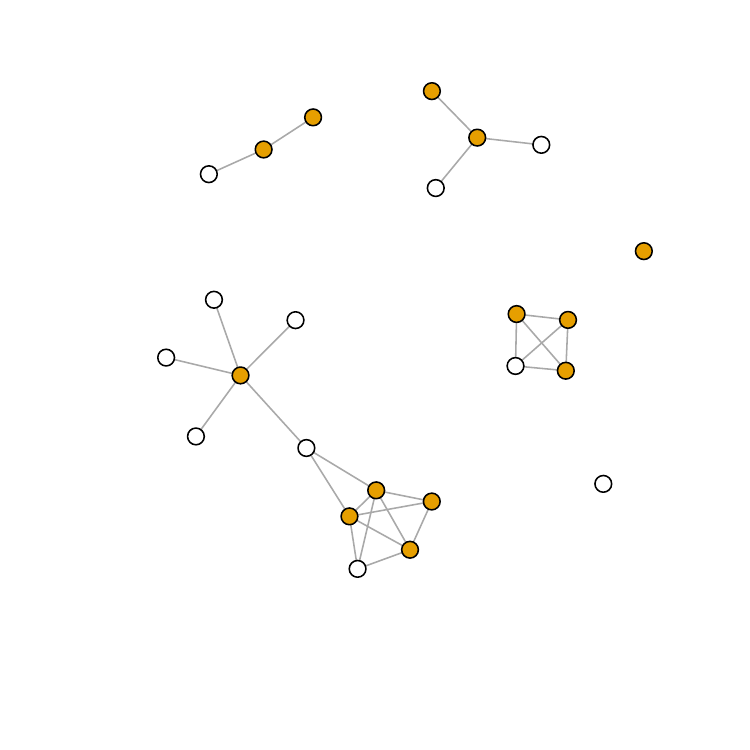}
\caption{End of term social network for one lab section of the whole-course ISLE curriculum.}
\label{fig:PostISLEFullLabNetwork}
\end{subfigure}

\caption{Pre and post-term social networks for the lab section of the whole-course implementation of ISLE. The network survey was distributed within the individual lab sections. Small groupings of students are visible, likely due to collaborative lab groups.}
\label{fig:Networks-ISLEwholecourseLab}
\end{figure*}

\begin{figure*}[htbp]

\begin{subfigure}[b]{0.45\linewidth}
\includegraphics[width=\linewidth]{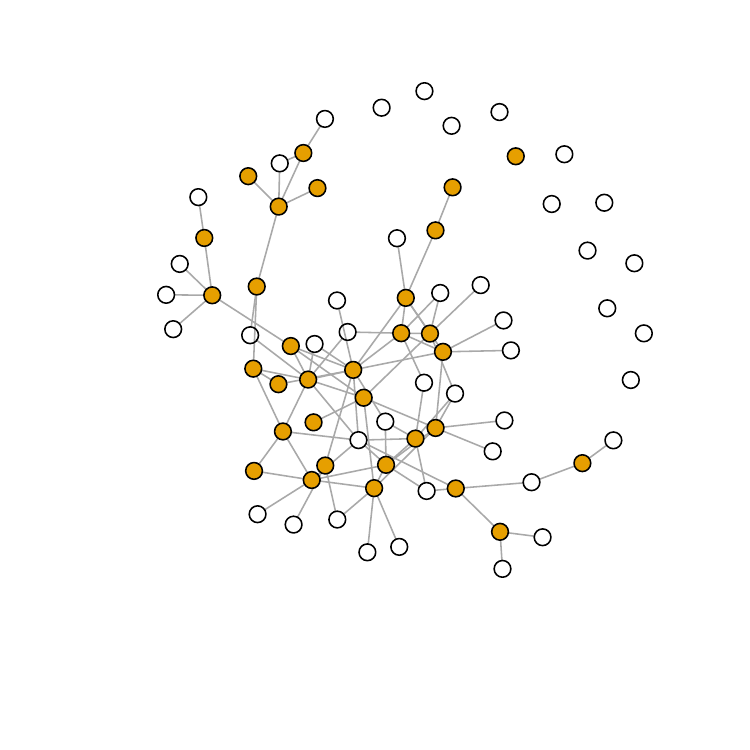} 
\caption{Beginning of term social network for one section of the Modeling Instruction curriculum.}
\label{fig:PreModelingNetwork}
\end{subfigure}
\begin{subfigure}[b]{0.45\linewidth}
\includegraphics[width=\linewidth]{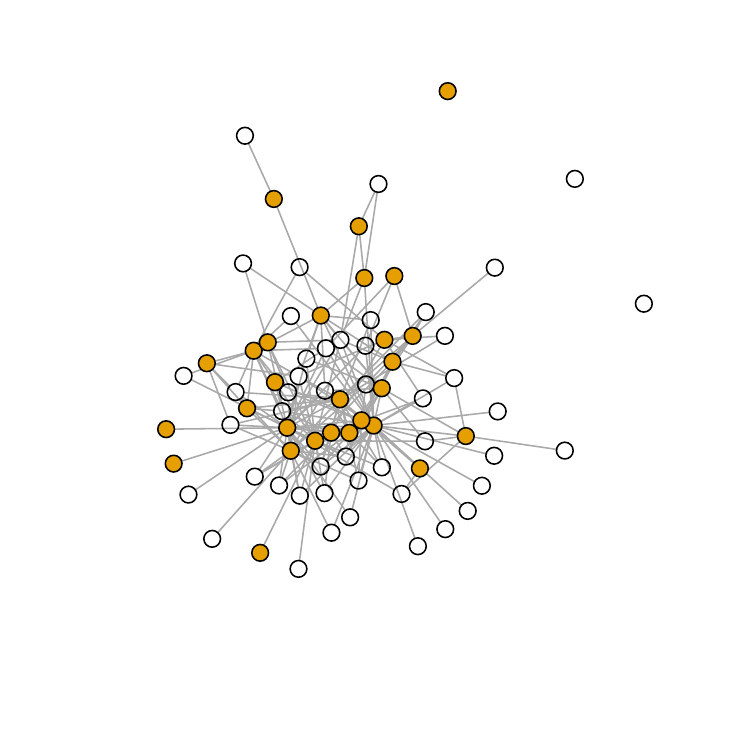}
\caption{End of term social network for one section of the Modeling Instruction curriculum.}
\label{fig:PostModelingNetwork}
\end{subfigure}

\caption{Pre and post-term social networks for Modeling Instruction. Network surveys were distributed by section. The pre-network is tightly connected, but still has two distinct islands at the top and right side of the network diagram. The post network shows a highly integrated class structure. The color shading on these plots indicates which students filled out the survey.}
\label{fig:Networks-Modeling}
\end{figure*}

\begin{figure*}[htbp]

\begin{subfigure}[b]{0.45\linewidth}
\includegraphics[width=\linewidth]{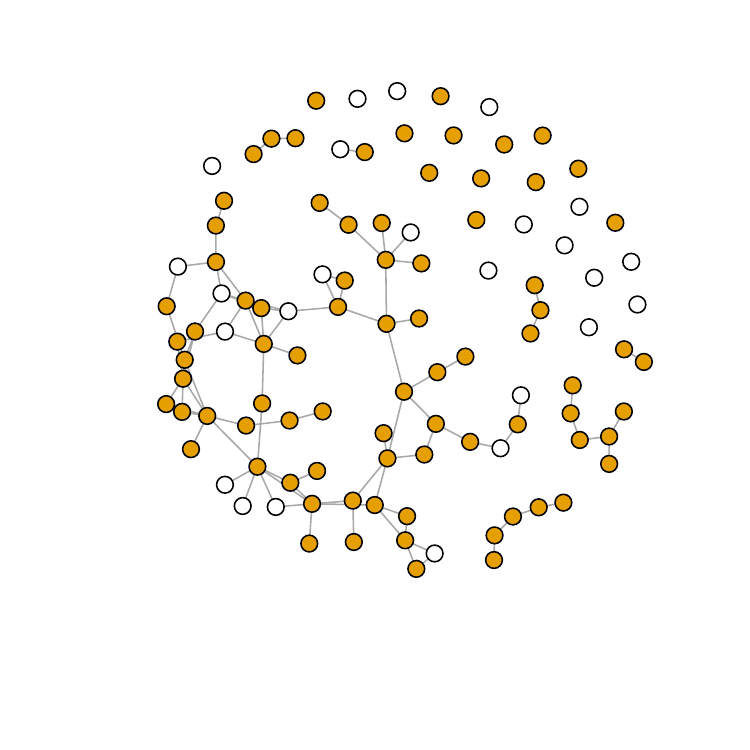} 
\caption{Beginning of term social network for one section of the Peer Instruction curriculum.}
\label{fig:PrePINetwork}
\end{subfigure}
\begin{subfigure}[b]{0.45\linewidth}
\includegraphics[width=\linewidth]{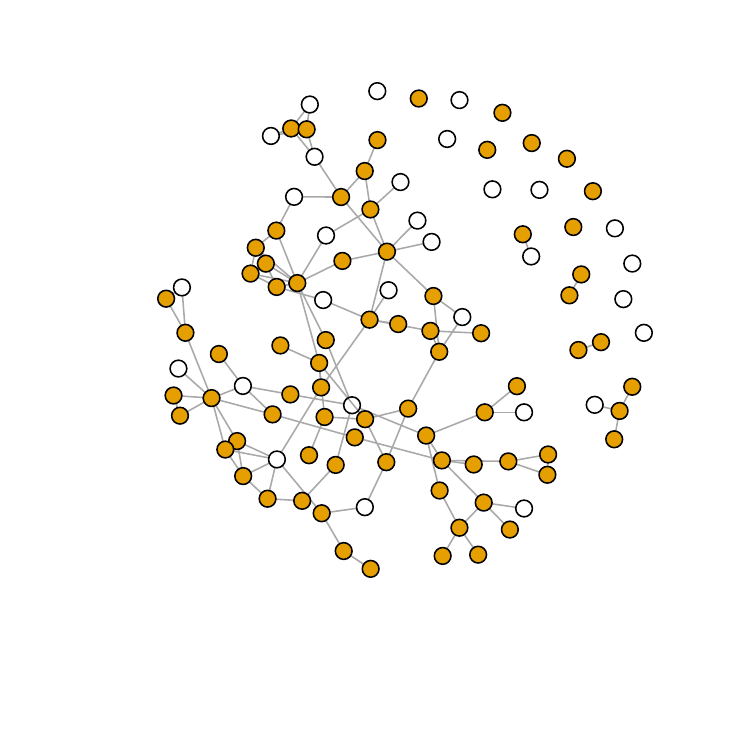}
\caption{End of term social network for one section of the Peer Instruction curriculum.}
\label{fig:PostPINetwork}
\end{subfigure}

\caption{Pre and post-term social networks for Peer Instruction. The network survey was distributed at the lecture level. Small pairings of students are visible, likely due to think-pair-share types of clicker questions. We also see a high number of isolates, likely due to the open-seating arrangement that does not require students to interact with their peers. Additionally, we see long chains of students, likely due to the layout of the lecture hall. The color shading on these plots indicates which students filled out the survey.}
\label{fig:Networks-PeerInstruction}
\end{figure*}

\begin{figure*}[htbp]

\begin{subfigure}[b]{0.45\linewidth}
\includegraphics[width=\linewidth]{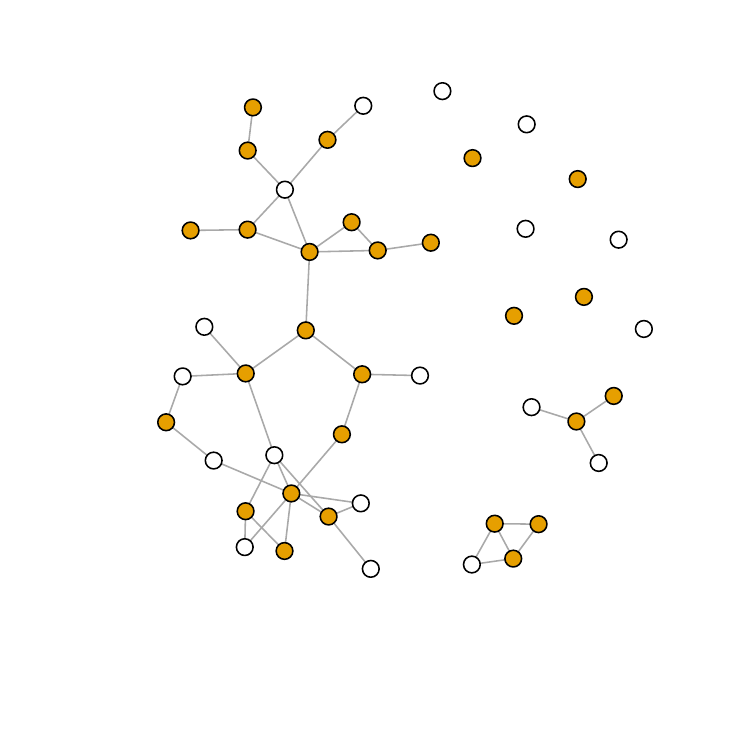} 
\caption{Beginning of term social network for one section of the Context-Rich Problems curriculum.}
\label{fig:PreCRPNetwork}
\end{subfigure}
\begin{subfigure}[b]{0.45\linewidth}
\includegraphics[width=\linewidth]{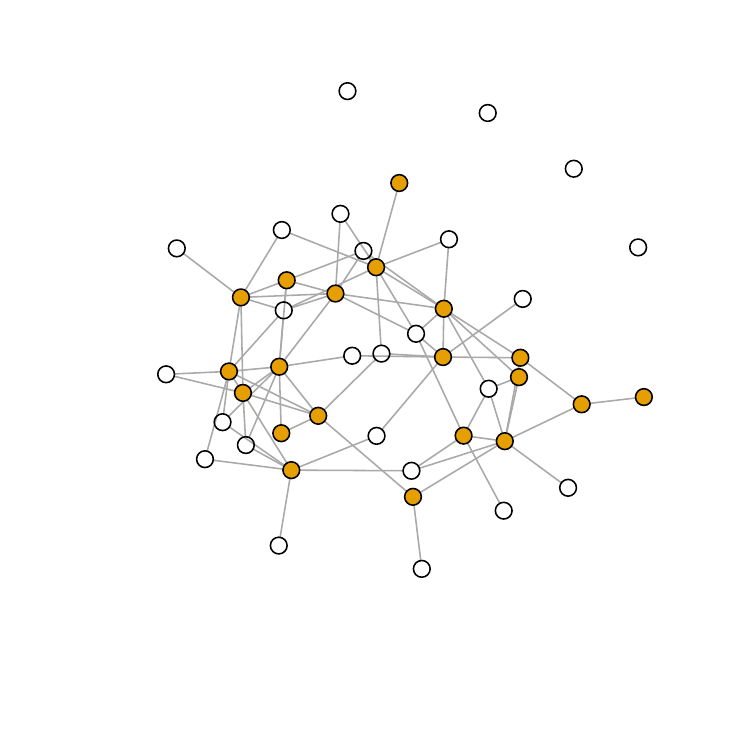}
\caption{End of term social network for one section of the Context-Rich Problems curriculum.}
\label{fig:PostCRPNetwork}
\end{subfigure}

\caption{Pre and post-term social networks for Context-Rich Problems. The network survey was distributed at the lecture/discussion level. Small groupings of students are visible in the pre network, likely due to discussion or lab groups. The post network is significantly more connected. The color shading on these plots indicates which students filled out the survey.}
\label{fig:Networks-CRP}
\end{figure*}

\begin{figure*}[htbp]

\begin{subfigure}[b]{0.45\linewidth}
\includegraphics[width=\linewidth]{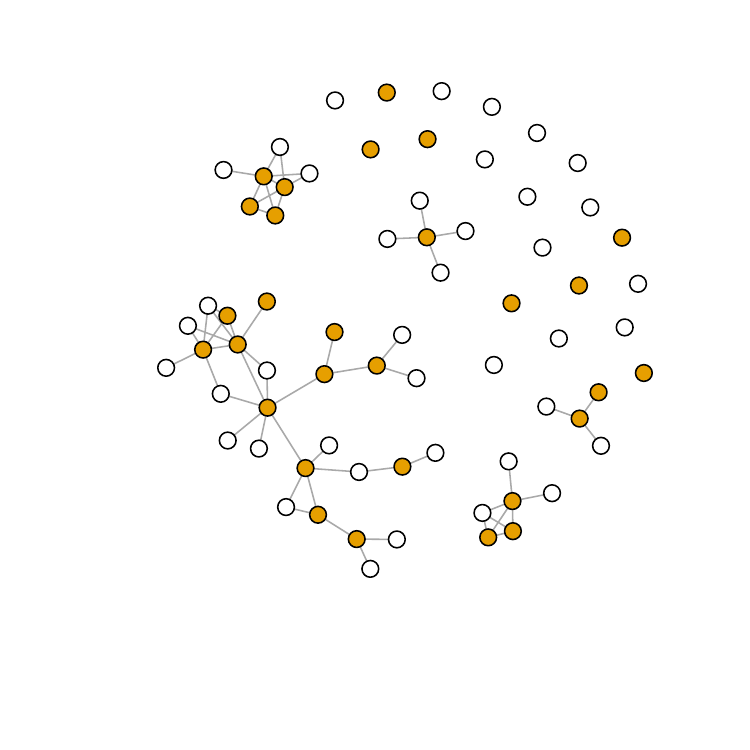} 
\caption{Beginning of term social network for the SCALE-UP curriculum.}
\label{fig:PreSCALEUPNetwork}
\end{subfigure}
\begin{subfigure}[b]{0.45\linewidth}
\includegraphics[width=\linewidth]{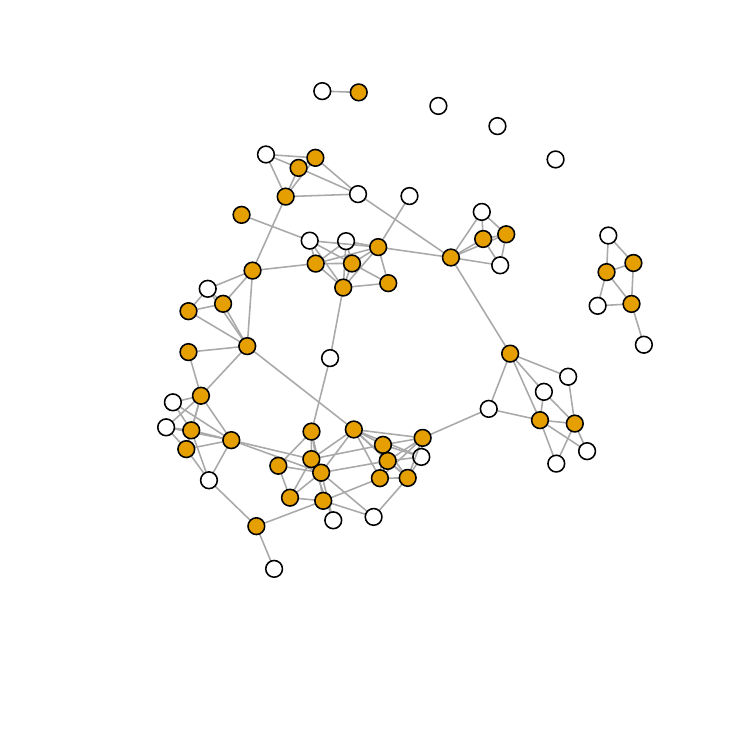}
\caption{End of term social network for the SCALE-UP curriculum.}
\label{fig:PostSCALEUPNetwork}
\end{subfigure}

\caption{Pre and post-term social networks for SCALE-UP. The network survey was distributed at the section level. Larger groupings of students are visible, likely due to the large table dynamic within the SCALE-UP room. The color shading on these plots indicates which students filled out the survey.}
\label{fig:Networks-SCALE-UP}
\end{figure*}

\bibliography{networks_classrooms}